\documentclass[prd,nopacs,preprintnumbers,twocolumn,amsmath,nofootinbib,amssymb]{revtex4}
\usepackage{graphicx,color,dcolumn,booktabs,bm}
\usepackage{longtable,lscape}
\usepackage{txfonts}
\usepackage{overpic}
\usepackage{epstopdf}
\usepackage{indentfirst}
\usepackage{feynmf}   
\usepackage{slashed}  
\usepackage{cases}
\usepackage{color}
\usepackage{float}
\usepackage{gensymb}
\usepackage{multirow}
\usepackage{epsfig,dsfont,amssymb,amsmath,amsfonts,amsbsy,mathrsfs}

\graphicspath{{}} %

\usepackage{hyperref}
\hypersetup{colorlinks,citecolor=blue,anchorcolor=red,menucolor=red,linkcolor=red,filecolor=red,runcolor=red,urlcolor=blue,frenchlinks=red}

\makeatletter
\@addtoreset{equation}{section}
\makeatother

\allowdisplaybreaks

\newcolumntype{I}{!{\vrule width 0.9pt}}

\begin{document}

\title{Hidden-bottom hadronic decays of $\Upsilon(10753)$ with a $\eta^{(\prime)}$ or $\omega$ emission}
\author{Yu-Shuai Li$^{1,2}$}\email{liysh20@lzu.edu.cn}
\author{Zi-Yue Bai$^{1,2}$}\email{baizy15@lzu.edu.cn}
\author{Qi Huang$^{3,4}$}\email{huangqi@ucas.ac.cn}
\author{Xiang Liu$^{1,2,4}$\footnote{Corresponding author}}\email{xiangliu@lzu.edu.cn}
\affiliation{$^1$School of Physical Science and Technology, Lanzhou University, Lanzhou 730000, China\\
$^2$Research Center for Hadron and CSR Physics, Lanzhou University and Institute of Modern Physics of CAS, Lanzhou 730000, China\\
$^3$University of Chinese Academy of Sciences (UCAS), Beijing 100049, China\\
$^4$Lanzhou Center for Theoretical Physics and Frontier Science Center for Rare Isotopes, Lanzhou University, Lanzhou 730000, China}

\begin{abstract}
In this work, we propose the $4S$-$3D$ mixing scheme to assign the $\Upsilon(10753)$ into the conventional bottomonium family. Under this interpretation, we further study its hidden-bottom hadronic decays with a $\eta^{(\prime)}$ or $\omega$ emission, which include $\Upsilon(10753)\to\Upsilon(1S)\eta^{(\prime)}$, $\Upsilon(10753)\to h_{b}(1P)\eta$ and $\Upsilon(10753)\to\chi_{bJ}\omega$ ($J$=0,1,2) processes. Since the $\Upsilon(10753)$ is above the $B\bar{B}$ threshold, the coupled-channel effect cannot be ignored, thus, when calculating partial decay widths of these $\Upsilon(10753)$ hidden-bottom decays, we apply the hadronic loop mechanism. Our result shows that these discussed decay processes own considerable branching fractions with the order of magnitude of $10^{-4}\sim 10^{-3}$, which can be accessible at Belle II and other future experiments.
\end{abstract}

\pacs{} %
\maketitle

\section{introduction}
\label{sec1}

When checking the status of the bottomonia collected in Particle Data Group (PDG) \cite{Zyla:2020zbs}, it is easy to find that compared with charmonium family, bottomonium family is far from been established. Until now, above the $B\bar{B}$ threshold, experiments have only observed three vector bottomonia $\Upsilon(10580)$, $\Upsilon(10860)$ and $\Upsilon(11020)$. Thus, to construct bottomonium family, finding more bottomonia is a crucial step. Along this line, the Belle and further Belle II experiments are playing important roles in exploring new bottomonium \cite{Kou:2018nap}.

Recently, the Belle Collaboration reanalyzed the cross section of $e^{+}e^{-}\to\Upsilon(nS)\pi^{+}\pi^{-}$ ($n=1,2,3$) and updated their measurements to supersede the previous result \cite{Santel:2015qga,Abdesselam:2019gth}. In their new analysis, apart from the already known resonances $\Upsilon(10860)$ and $\Upsilon(11020)$ \cite{Besson:1984bd,Lovelock:1985nb,Santel:2015qga,He:2014sqj,Abdesselam:2015zza,Yin:2018ojs}, a new structure near 10.75 GeV, which is refereed to be $\Upsilon(10753)$ in PDG \cite{Zyla:2020zbs}, appears in the $\Upsilon(nS)\pi^{+}\pi^{-}$ ($n=1,2,3$) invariant mass spectrum, whose mass and width are fitted as $M=(10752.7\pm5.9^{+0.7}_{-1.1})$ MeV and $\Gamma=(35.5^{+17.6 +3.9}_{-11.3 -3.3})$ MeV, respectively. Additionally, its spin-parity is indicated to be $J^{PC}=1^{--}$ by the requirement of the production processes.
This new reported bottomonium-like structure quickly inspired theorist's interests in revealing its inner structure, where
the conventional bottomonium state assignment \cite{Chen:2019uzm,Li:2019qsg} and several exotic state interpretations, which include tetraquark state \cite{Ali:2020svd,Wang:2019veq} and hybrid state \cite{TarrusCastella:2019lyq,TarrusCastella:2021fob}, and kinetic effect \cite{Bicudo:2019ymo,Bicudo:2020qhp} were proposed.

When treating the $\Upsilon(10753)$ as a conventional bottomonium state, we have to face a serious problem. The calculation of the mass spectrum of bottomonium family \cite{Segovia:2016xqb,Wang:2018rjg,Godfrey:2015dia,Badalian:2008ik,Badalian:2009bu} shows that the predicted masses of $\Upsilon(4S)$, $\Upsilon(3D)$, $\Upsilon(5S)$ and $\Upsilon(4D)$ are around $(10607 - 10640)$ MeV, $(10653 - 10717)$ MeV, $(10818 - 10878)$ MeV and $(10853 - 10928)$ MeV, respectively. Obviously, the mass of the observed $\Upsilon(10753)$ can not fall into these predicted mass ranges, which is the difficulty to assign the $\Upsilon(10753)$ to be a bottomonium state. Naturally, the exotic state explanations \cite{Ali:2020svd,Wang:2019veq,TarrusCastella:2019lyq,TarrusCastella:2021fob} were given since this mass problem can be solved by this way.

In fact, we should mention some experiences on constructing the charmonium family.
{In Ref. \cite{Rosner:2001nm}, in order to understand the puzzling phenomenon involved in the $\psi(3686)$ and $\psi(3770)$, Rosner introduced the $2S$ and $1D$ state mixing scheme.}
{Furthermore, when checking the higher charmonia, such effect also should be emphasized.}
In Ref. \cite{Wang:2019mhs}, the predicted masses of $\psi(4S)$ and $\psi(3D)$ are 4274 MeV and 4334 MeV respectively. If assigning the $Y(4220)$ \cite{Zyla:2020zbs} as a $\psi(4S)$ charmonium state, the mass of $\psi(4S)$ is about 54 MeV higher than the mass of $Y(4220)$ \cite{Wang:2019mhs}. To solve this problem, the $4S$-$3D$ mixing scheme is introduced in Ref. \cite{Wang:2019mhs}. It is found that when the mixing angle is around $(33 \pm 3)\degree$, the mass of $Y(4220)$ can be reproduced, simultaneously, a partner state has mass of 4380 MeV \cite{Wang:2019mhs}.
{Moreover, the study around these charmoniumlike states around 4.6 GeV also shows the important role of the $S$-$D$ mixing scheme when putting them into the charmonium family \cite{Wang:2020prx}.}
{So from these previous experience on constructing the charmonium family, we may find that the $S$-$D$ mixing effect is universal and cannot be ignored when depicting heavy quarkonium spectroscopy.}

Inspired by the research experiences of constructing the charmonium family \cite{Rosner:2001nm,Wang:2019mhs,Wang:2020prx}, in this work we suggest that introducing the $S$-$D$ mixing scheme can also solve the mass problem of the $\Upsilon(10753)$. Firstly, checking PDG data \cite{Zyla:2020zbs}, one notices that the experimental mass of $\Upsilon(10580)$ ($10579.4$ MeV) is lower than the prediction of $\Upsilon(4S)$ ($10607-10640$ MeV) \cite{Segovia:2016xqb,Wang:2018rjg,Godfrey:2015dia,Badalian:2008ik,Badalian:2009bu}. And, the mass of the newly reported $\Upsilon(10753)$ is higher than the predicted mass of $\Upsilon(3D)$  ($10653-10717$ MeV). This situation satisfies the requirement of introducing the $4S$-$3D$ mixing, by which the mass problem of the $\Upsilon(10753)$ can be solved. In Sec. \ref{sec2}, we will present the details of the introduced $4S$-$3D$ mixing scheme.

It is obvious that the bottomonium assignment to the $\Upsilon(10753)$
can survive. Thus, how to distinguish these several explanations on $\Upsilon(10753)$ becomes a key point for now, which means studies on the decay behaviors of $\Upsilon(10753)$ under different assignments should be paid more attention.
Till now, we notice that although there exists theoretical estimate on the corresponding open-bottom hadronic decay modes \cite{Liang:2019geg}, the study of its hidden-bottom hadronic decays is still absent. This status stimulates our interest in investigating the hidden-bottom hadronic decays of $\Upsilon(10753)$ with the bottomonium assignment.

When looking at PDG \cite{Zyla:2020zbs}, it is easy to find that for some hidden-bottom hadronic decays of higher bottomonia, their decay rates are anomalous \cite{Zyla:2020zbs}.
For example, even for some spin flip transitions forbidden by heavy quark spin symmetry, their decay rates are still considerable \cite{Voloshin:2007dx,Anwar:2017urj,Voloshin:2012dk}.
Since these higher bottomonia mainly decay into a pair of $B$ mesons, it is easy to estimate that coupled-channel effect may be important.
Thus, as an equivalent description of the coupled-channel effect, the hadronic loop mechanism was introduced to understand the puzzling phenomena involved in hidden-bottom hadronic decays of some higher bottomonia \cite{Meng:2008bq,Zhang:2018eeo,Wang:2016qmz,Chen:2014ccr,Chen:2011jp,Huang:2018pmk,Huang:2018cco,Huang:2017kkg,Meng:2007tk,Meng:2008dd,Chen:2011qx,Chen:2011zv}. For example, after introducing the contribution from the hadronic loop mechanism, the anomalous branching fractions such as $\Upsilon(10580)\to h_{b}(1P)\eta$ \cite{Tamponi:2015xzb}, $\Upsilon(10580)\to\Upsilon(nS)\pi^+\pi^-$ \cite{Guido:2017cts}, $\Upsilon(10580)\to\Upsilon(1S)\eta$ \cite{Guido:2017cts}, $\Upsilon(10860)\to\Upsilon(nS)\pi^+\pi^-$ \cite{Abdesselam:2019gth}, $\Upsilon(10860)\to h_b(nP)\pi^+\pi^-$ \cite{Adachi:2011ji}, $\Upsilon(10860,11020)\to\pi^+\pi^-\pi^0\chi_{bJ}$ \cite{He:2014sqj,Yin:2018ojs} can be well understood \cite{Meng:2008bq,Zhang:2018eeo,Wang:2016qmz,Chen:2014ccr,Chen:2011jp,Huang:2018pmk,Huang:2018cco,Huang:2017kkg,Meng:2007tk,Meng:2008dd,Chen:2011qx,Chen:2011zv}. Since $\Upsilon(10753)$ is also above the $B\bar{B}$ threshold, borrowing the former experience of studying these higher bottomonia like $\Upsilon(10580)$, $\Upsilon(10860)$ and $\Upsilon(11020)$
\cite{Meng:2008bq,Zhang:2018eeo,Wang:2016qmz,Chen:2014ccr,Chen:2011jp,Huang:2018pmk,Huang:2018cco,Huang:2017kkg,Meng:2007tk,Meng:2008dd,Chen:2011qx,Chen:2011zv}, the hadronic loop mechanism cannot be ignored when calculating the hidden-bottom hadronic decays of $\Upsilon(10753)$ .

In this work, by taking into account the hadronic loop mechanism, we perform a study on the hidden-bottom decay channels $\Upsilon(10753)\to \Upsilon(1S)\eta^{(\prime)}$, $h_{b}(1P)\eta$ and $\chi_{bJ}\omega$, where the corresponding partial decay widths and branching fractions are estimated. Our results must be an important part of the physics around the newly reported $\Upsilon(10753)$, and can provide useful information to the forthcoming experiments when exploring these processes.
With the joint efforts from theorists and experimentalists, we have reasons to believe that the feature of the discussed $\Upsilon(10753)$ can be understood well.

This paper is organized as follows.
After the introduction, in Sec. \ref{sec2} we introduce the $4S$-$3D$ mixing mechanism to assign the $\Upsilon(10753)$ as a conventional bottomonium state. Then, we illustrate the detailed calculation of $\Upsilon(10753)\to\Upsilon(1S)\eta^{(\prime)}$, $h_{b}(1P)\eta$ and $\chi_{bJ}\omega$ in Sec. \ref{sec3}.
After that, we present numerical results in Sec. \ref{sec4}.
Finally, the paper ends with a short summary.

\section{The $4S$-$3D$ mixing scheme}
\label{sec2}

Although there exists exotic state \cite{Ali:2020svd,Wang:2019veq,TarrusCastella:2019lyq,TarrusCastella:2021fob} and kinetic effect \cite{Bicudo:2019ymo,Bicudo:2020qhp} interpretations to the newly observed $\Upsilon(10753)$, we still need to meticulously study the possibility that if this $\Upsilon(10753)$ can be included into the bottomonium family. In Ref. \cite{Wang:2018rjg}, a systematical investigation on the higher bottomonia was carried out by using a modified Godfrey-Isgur model\footnote{{As an unquenched quark model, the modified Godfrey-Isgur model is an equivalent description of the coupled-channel effect \cite{Wang:2018rjg} since the screening potential was introduced \cite{Ding:1995he,Li:2009ad}. 
With the modified Godfrey-Isgur model, the mass of $\Upsilon(3^3D_1)$ state was predicted \cite{Wang:2018rjg}. In this work, we adopt this value as input. 
We noticed that there exists a systematical study of bottomonium meson family with a realistic coupled-channel calculation by considering 
these open-bottom meson pair channels \cite{Liu:2011yp}, where the predicted mass of $\Upsilon(3^3D_1)$ is 10650.9 MeV, which is comparable with our adopted result $10675$ MeV taken from Ref. \cite{Liu:2011yp}. As emphasized in Refs. \cite{Liu:2011yp,Wang:2018rjg}, the coupled-channel effect is obvious since the mass shift between bare mass and physical mass for $\Upsilon(3^3D_1)$ is around 90 MeV. This situation inspires us to introduce hadronic loop contribution when estimating the branching ratios of the $\Upsilon(3^3D_1)$ decays into hidden-charm decay channels in this work.}} and predicted mass and width of the $\Upsilon(3D)$ state are 10675 MeV and 54.1 MeV, respectively.

Although the predicted width of $\Upsilon(3D)$ is consistent with the current measurement of the $\Upsilon(10753)$ \cite{Abdesselam:2019gth}, the predicted mass is about 70 MeV lower than the measurement. Furthermore, after checking other theoretical results on the mass spectrum of bottomonium family, we find that the predicted mass of $\Upsilon(3D)$ is in the range $(10653 - 10717)$ MeV \cite{Segovia:2016xqb,Wang:2018rjg,Godfrey:2015dia,Badalian:2008ik,Badalian:2009bu}, which also shows that none of the predicted mass of $\Upsilon(3D)$ is close to the current measurement. In addition, after checking PDG \cite{Zyla:2020zbs}, we find that the mass of $\Upsilon(10580)$ is 10579.4 MeV, which is also far away from the predicted mass $(10607-10640)$ MeV \cite{Segovia:2016xqb,Wang:2018rjg,Godfrey:2015dia,Badalian:2008ik,Badalian:2009bu}.

Since the experimental masses of the $\Upsilon(10753)$ and $\Upsilon(10580)$ is higher and lower than the corresponding theoretical results, which stimulate the idea to introduce the $S$-$D$ mixing mechanism, where this mixing mechanism had been successfully applied to understand the observed charmonium-like states \cite{Rosner:2001nm,Wang:2019mhs,Wang:2020prx,Qian:2021gby}. Especially, in Ref. \cite{Wang:2019mhs}, after introducing the $4S$-$3D$ mixing, the mass of $Y(4220)$, whose mass is about 54 MeV lower than prediction, is reproduced well. And its partner $\psi(4380)$ was predicted, which has showed some hints in the $e^+ e^- \to D \bar{D}^\ast_2 \to D^0 D^- \pi^+$, $e^+ e^- \to D^0 D^{\ast-} \pi^+$, and $e^+ e^- \to \psi(2S) \pi^+ \pi^-$ processes \cite{Wang:2019mhs}.

Thus, in our view it is meaningful to investigate whether or not the introduction of $4S$-$3D$ mixing can solve the mass problem of the $\Upsilon(10753)$. Similar to the approach in Refs. \cite{Rosner:2001nm,Wang:2019mhs,Wang:2020prx}, we introduce
\begin{equation}
\left(\begin{array}{c}
\Upsilon^{\prime}_{4S}\\
\Upsilon^{\prime}_{3D}
\end{array}\right)
=\left(\begin{array}{cc}
\cos{\theta} & -\sin{\theta}\\
\sin{\theta} & \cos{\theta}
\end{array}\right)
\left(\begin{array}{c}
\Upsilon_{4S}\\
\Upsilon_{3D}
\end{array}\right),
\end{equation}
to describe the 4$S$-3$D$ mixing, where $\theta$ denotes the mixing angle. $\Upsilon_{4S}$ and $\Upsilon_{3D}$ are the wave functions of the pure $4S$ and $3D$ bottomonium states, respectively. $\Upsilon_{4S}^\prime$ and $\Upsilon_{3D}^\prime$ are the wave functions of the physical states, respectively.

To fix the mixing angle $\theta$, in this work we focus on the electronic decay of bottomonium states since we already have the experimental data of $\Gamma_{ee}(\Upsilon(10580))$ \cite{Zyla:2020zbs}. The expressions we used to fit the electronic decay width are as follows \cite{Li:2009nr}
\begin{equation}
\begin{split}
\Gamma_{ee}(nS)&=\frac{4\alpha^2e_b^2}{M_{nS}^2}|R_{nS}(0)|^2\left(1-\frac{16}{3}\frac{\alpha_{s}}{\pi}\right),\\
\Gamma_{ee}(nD)&=\frac{4\alpha^2e_b^2}{M_{nD}^2}|\frac{5}{2\sqrt{2}m_b^2}R^{\prime\prime}_{nD}(0)|^2\left(1-\frac{16}{3}\frac{\alpha_{s}}{\pi}\right),
\label{dielectronic width}
\end{split}
\end{equation}
where $e_b=1/3$ is the charge of $b$ quark, $\alpha$ is the fine structure constant and $\alpha_{s}=0.18$ \cite{Wang:2018rjg}.
Besides, $R_{nS}(0)$ is the radial $S$ wave function at the origin, while $R_{nD}^{\prime\prime}(0)$ is the second derivative of the radial $D$ wave function at the origin. After considering the $4S-3D$ mixing, the electronic decay width of the mixed $\Upsilon^{\prime}(4S)$ state is
\begin{equation}
\begin{split}
\Gamma_{ee}=&\frac{4\alpha^2e_b^2}{M_{\Upsilon^{\prime}(4S)}^2}\left |R_{nS}(0)\cos{\theta}+\frac{5}{2\sqrt{2}m_b^2}R^{\prime\prime}_{nD}(0)\sin{\theta}\right |^2\\
&\times\left(1-\frac{16}{3}\frac{\alpha_{s}}{\pi}\right).\label{eq:4S-ee}
\end{split}
\end{equation}

With Eq. (\ref{eq:4S-ee}) and the experimental data of the electronic decay width of the $\Upsilon(10580)$ ($\Gamma_{ee}(\Upsilon(10580)) = (0.272\pm0.029)\ \text{KeV}$)  \cite{Zyla:2020zbs}, and after substituting into the $R_{nS}(0)$ and $R^{\prime\prime}_{nD}(0)$ extracted from Ref. \cite{Wang:2018rjg}, the mixing angle is fixed to be $\theta=(33\pm4)\degree$, which is close to the results given in Refs. \cite{Badalian:2008ik,Badalian:2009bu}.

\begin{figure}[htbp]\centering
  \includegraphics[width=70mm]{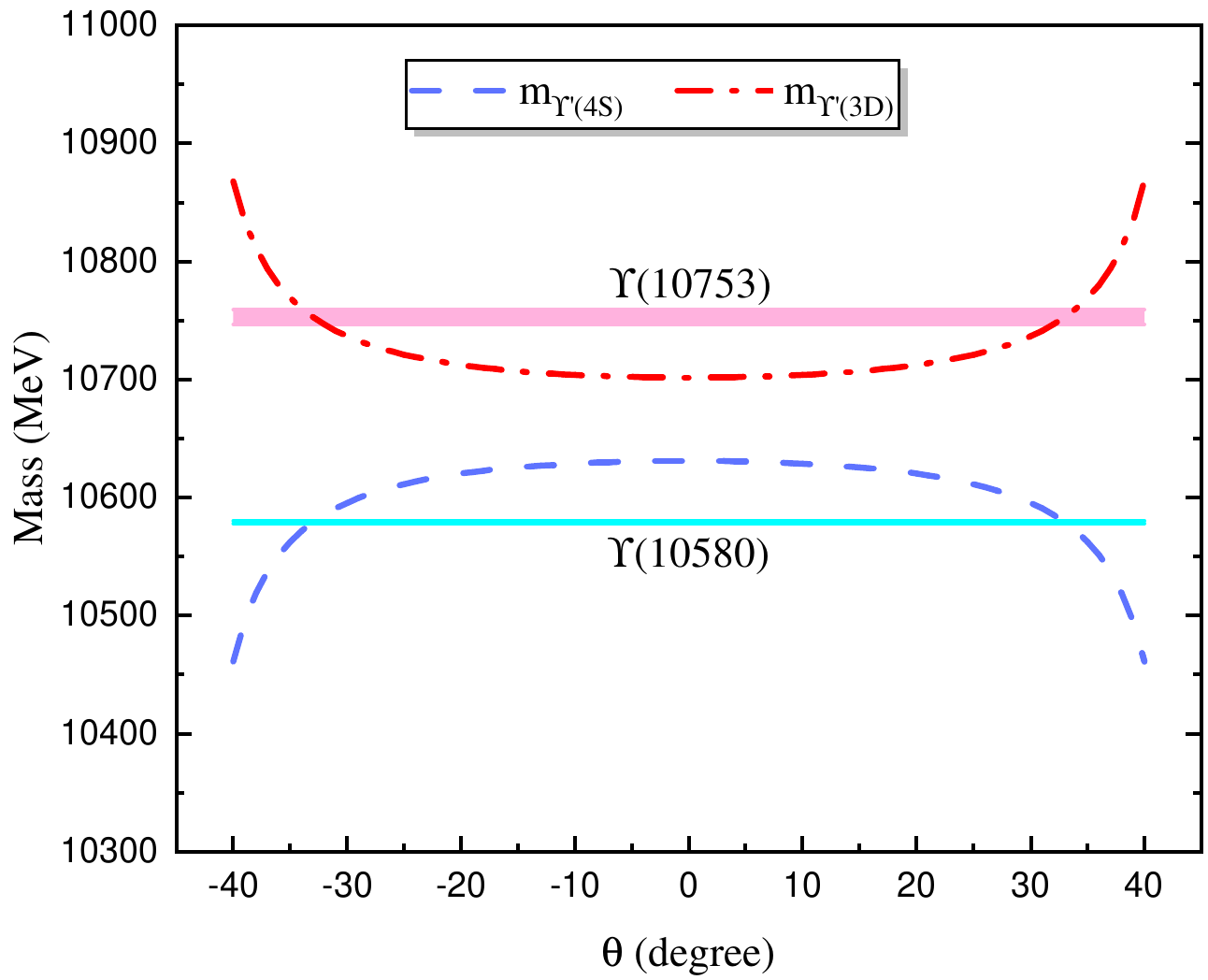}\\
  \caption{The $\theta$ dependence of the masses of the physical states $\Upsilon^{\prime}(4S)$ and $\Upsilon^{\prime}(3D)$. Here, the cyan and pink bands denote the current measurements of $m_{\Upsilon(10580)}$ and $m_{\Upsilon(10753)}$, respectively. The masses of pure state $\Upsilon(4S)$ and $\Upsilon(3D)$ is taken as the range of $(10621 - 10643)\ \text{MeV}$ and $(10691 - 10712)\ \text{MeV}$, respectively, which are from theoretical predictions \cite{Segovia:2016xqb,Wang:2018rjg,Godfrey:2015dia,Badalian:2008ik,Badalian:2009bu}.}
\label{mixingtheta}
\end{figure}

Then, we turn to illustrate the change of mass caused by the mixing. In this mixing scheme, the mass eigenvalues of the physical states $\Upsilon^{\prime}(4S)$ and $\Upsilon^{\prime}(3D)$ are determined by
\begin{equation}
\begin{split}
m^2_{\Upsilon^{\prime}(4S)}&=\frac{1}{2}\left [m^2_{\Upsilon(4S)}+m^2_{\Upsilon(3D)}-\sqrt{(m^2_{\Upsilon(4S)}-m^2_{\Upsilon(3D)})^2 \text{sec}^2 2\theta}\right ],\\
m^2_{\Upsilon^{\prime}(3D)}&=\frac{1}{2}\left[ m^2_{\Upsilon(4S)}+m^2_{\Upsilon(3D)}+\sqrt{(m^2_{\Upsilon(4S)}-m^2_{\Upsilon(3D)})^2 \text{sec}^2 2\theta}\right],\label{eq:mass-4S-3D-mixing}
\end{split}
\end{equation}
where $m_{\Upsilon(4S)}$ and $m_{\Upsilon(3D)}$ are the masses of pure $4S$ and $3D$ bottomonium states, respectively.
Based on the Eq. (\ref{eq:mass-4S-3D-mixing}), we investigate the $\theta$-dependence of $m_{\Upsilon^{\prime}(4S)}$ and $m_{\Upsilon^{\prime}(3D)}$.
As shown in Fig. \ref{mixingtheta}, the mass of the physical state $\Upsilon^{\prime}(4S)$ is decreased while the mass of the physical state $\Upsilon^{\prime}(3D)$ is increased when increasing the mixing angle $|\theta|$.
Thus, introducing the $4S$-$3D$ mixing scheme can reproduce the mass of the $\Upsilon(10753)$ and $\Upsilon(10580)$ well.

{In this section, we discuss the $S$-$D$ mixing scheme for $\Upsilon(10750)$ and $\Upsilon(10580)$. In fact, we have reason to believe that there should exist the mixing scheme for other pairs of quantum numbers and these mixing schemes would modify the entire spectrum. However, the present experimental information of heavy quarkonium especially for bottomonium is not enough to be applied to test such scenario. For example, $F$-wave charmonium states are still absent, which may have mixing with the corresponding $P$-wave state. For bottomonium, the $\Upsilon(1^3D_1)$ and $\Upsilon(2^3D_1)$ states are still missing in experiment. We hope that this situation can be changed with the running of Belle II. Obviously, how the mixing scheme mentioned above changes the spectroscopy behavior of heavy quarkonium is an interesting research topic not only at present but also in the future.}

In general, by considering the $4S$-$3D$ mixIng, we find a way to resolve the mass puzzle of the $\Upsilon(10580)$ and $\Upsilon(10753)$ simultaneously, where the $\Upsilon(10753)$ can be assigned as the mixture of the $4S$ and $3D$ bottomonium states. In the following, we adopt this mixing scheme to present the corresponding hidden-bottom decay behavior.

\section{the transitions via the hadronic loop mechanism}
\label{sec3}

According to the latest Belle's result, the $\Upsilon(10753)$ is above $B^{(*)}\bar{B}^{(*)}$ and $B_{s}\bar{B}_{s}$ thresholds. Thus, the $\Upsilon(10753)$ should dominantly decay into a pair of bottomed or bottom-strange mesons.
Under the framework of hadronic loop mechanism \cite{Zhang:2018eeo,Meng:2008bq,Chen:2014ccr,Wang:2016qmz,Huang:2018pmk,Huang:2018cco,Huang:2017kkg}, the hidden-bottom hadronic decays $\Upsilon(10753)\to\Upsilon(1S)\eta^{(\prime)}$, $\Upsilon(10753)\to h_{b}(1P)\eta$ and $\Upsilon(10753)\to\chi_{bJ}\omega$ can be achieved by the following way.
Firstly, the initial $\Upsilon(10753)$ couples with $B^{(*)}\bar{B}^{(*)}$, then, the bottom meson pair converts into a low-lying bottomonia and one light meson by exchanging a $B^{(*)}$ meson. We should emphasize here that since the coupling between the $\Upsilon(10753)$ and $B_{s}\bar{B}_{s}$ is weak \cite{Liang:2019geg}, in this work we ignore the contributions from the intermediate $B_{s}$ meson loops.

\begin{figure}[htbp]\centering
  \includegraphics[width=80mm]{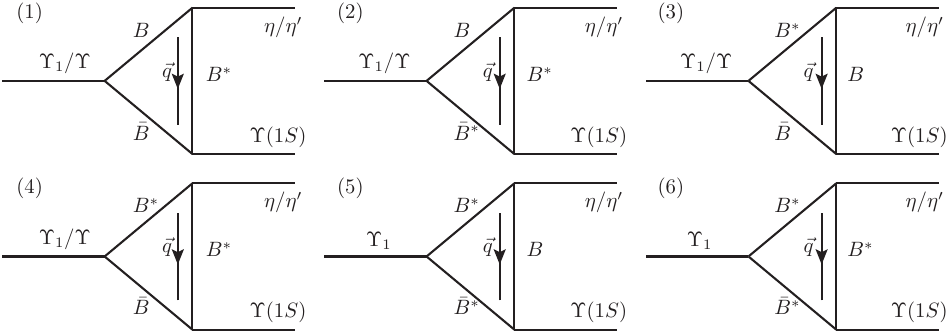}\\
  \caption{The schematic diagrams for depicting the $\Upsilon(4S,3D)\to\Upsilon(1S)\eta^{(\prime)}$ decay via the hadronic loop mechanism.}
\label{transitionUpseta}
\end{figure}

\begin{figure}[htbp]\centering
  \includegraphics[width=80mm]{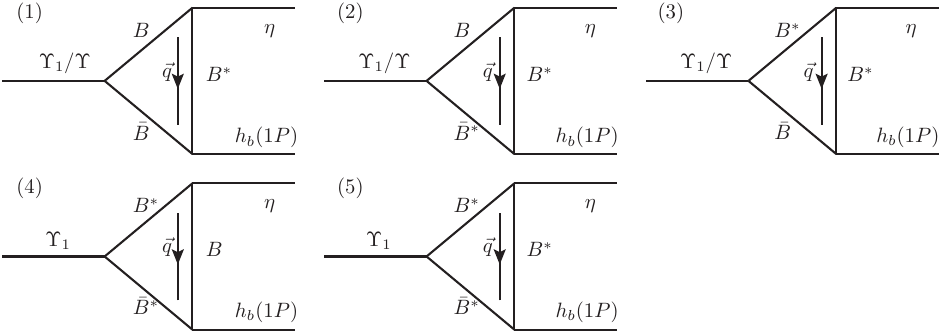}\\
  \caption{The schematic diagrams for depicting the $\Upsilon(4S,3D)\to h_{b}(1P)\eta$ decay via the hadronic loop mechanism.}
\label{transitionhbeta}
\end{figure}

\begin{figure}[htbp]\centering
  \includegraphics[width=80mm]{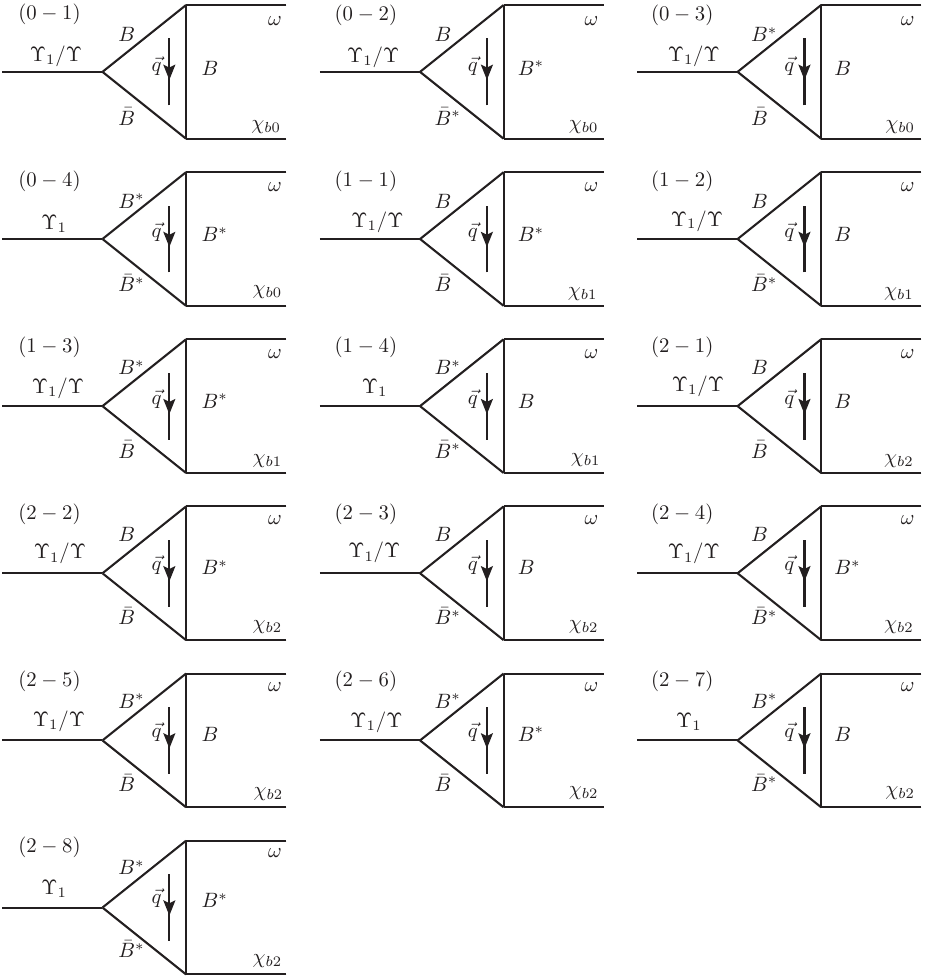}\\
  \caption{The schematic diagrams for depicting the $\Upsilon(4S,3D)\to\chi_{bJ}\omega$ ($J$=0,1,2) decay via the hadronic loop mechanism.}
\label{transitionchibJomega}
\end{figure}

According to the limits of quantum numbers and phase space, the diagrams depicting the involved decays can be determined.
As discussed in Sec. \ref{sec2}, the $\Upsilon(10753)$ can be well understood as the dressed $\Upsilon(3D)$ state with mixing the $\Upsilon(4S)$ component, so when we calculate the decay properties of the $\Upsilon(10753)$, we must consider not only the contributions from $3D$ component, but also the ones from $4S$ component.
The diagrams depicting the decays to $\Upsilon(1S)\eta^{(\prime)}$, $h_{b}(1P)\eta$ and $\chi_{bJ}(1P)\omega$ are shown in Fig. \ref{transitionUpseta}, Fig. \ref{transitionhbeta} and Fig. \ref{transitionchibJomega}, respectively.

The general expression of the amplitude can be written as
\begin{equation}
\mathcal{M}=\int\frac{d^{4}q}{(2\pi)^{4}}\frac{\mathcal{V}_{1}\mathcal{V}_{2}\mathcal{V}_{3}}{\mathcal{P}_{1}\mathcal{P}_{2}\mathcal{P}_{E}}\mathcal{F}^{2}(q^2,m_{E}^2),
\label{amplitude}
\end{equation}
where $\mathcal{V}_{\text{i}}$ (i=1,2,3) are interaction vertices, $1/\mathcal{P}_{1,2}$ and $1/\mathcal{P}_{E}$ denote the propagators of intermediate bottomed mesons. Here, $\mathcal{F}(q^{2},m_{E}^2)$ is the form factor, which is introduced to compensate the off-shell effect of the exchanged $B^{(*)}$ meson and depict the structure effect of interaction vertices \cite{Locher:1993cc,Li:1996yn,Cheng:2004ru}.
In this paper, the monopole form factor \cite{Gortchakov:1995im} is adopted, which is written as
\begin{equation}
\mathcal{F}(q^2,m_{E}^2)=\frac{\Lambda^2-m_{E}^2}{\Lambda^2-q^2},
\label{formfactor}
\end{equation}
where $m_{E}$ and $q$ denote the mass and momentum of the exchanged bottomed meson, respectively, and $\Lambda$ is the cutoff, which can be parameterized as $\Lambda=m_{E}+\alpha_{\Lambda}\Lambda_{QCD}$ with $\Lambda_{QCD}=220$ MeV \cite{Liu:2006dq,Liu:2009dr,Li:2013zcr} and $\alpha_{\Lambda}$ being a free parameter. Usually, the free parameter $\alpha_{\Lambda}$ should be of order one and is dependent on the concrete processes \cite{Cheng:2004ru}.

In order to write down the amplitudes, we adopt the effective lagrangian approach to describe the involved vertices $\mathcal{V}_{\text{i}}$ in Eq. (\ref{amplitude}).
Considering the heavy quark limit and chiral symmetry, the effective Lagrangians describing the interactions involved in bottomonium, bottomed (or bottom-strange) meson pair, and light pseudoscalar (vector) meson are \cite{Casalbuoni:1996pg,Wise:1992hn,Xu:2016kbn,Duan:2021bna}
\begin{equation}
\begin{split}
\mathcal{L}_{S}=&\ ig_{S}\text{Tr}[S^{(Q\bar{Q})}\bar{H}^{(\bar{Q}q)}\gamma^{\mu}\overset\leftrightarrow{\partial}_{\mu}\bar{H}^{(Q\bar{q})}]+\text{H.c.},\\
\mathcal{L}_{P}=&\
ig_{P}\text{Tr}[P^{(Q\bar{Q})\mu}\bar{H}^{(\bar{Q}q)}\gamma^{\mu}\bar{H}^{(Q\bar{q})}]+\text{H.c.},\\
\mathcal{L}_{D}=&\ ig_{D}\text{Tr}[D_{\mu\nu}^{(Q\bar{Q})}\bar{H}^{(\bar{Q}q)}\overset\leftrightarrow{\partial}_{\mu}\gamma^{\nu}\bar{H}^{(Q\bar{q})}]+\text{H.c.},\\
\mathcal{L}_{\mathbb{P}}=&\
ig_{H}\text{Tr}[H^{(Q\bar{q})\text{j}}\gamma_{\mu}\gamma_{5}(\mathcal{A}^{\mu})^{\text{i}}_{\text{j}}\bar{H}_{\text{i}}^{(Q\bar{q})}],\\
\mathcal{L}_{\mathbb{V}}=&\
i\beta\text{Tr}[H^{(Q\bar{q})\text{j}}\upsilon^{\mu}(-\rho_{\mu})^{\text{i}}_{\text{j}}\bar{H}_{\text{i}}^{(Q\bar{q})}]\\
&\ +i\lambda\text{Tr}[H^{(Q\bar{q})\text{j}}\sigma^{\mu\nu}F_{\mu\nu}(\rho^{\text{i}}_{\text{j}})\bar{H}_{\text{i}}^{(Q\bar{q})}]
\label{totalLagrangians}
\end{split}
\end{equation}
with $\overset\leftrightarrow{\partial}=\overset\rightarrow{\partial}-\overset\leftarrow{\partial}$. In Eq. \eqref{totalLagrangians}, $S^{(Q\bar{Q})}$, $P^{(Q\bar{Q})\mu}$ and $D_{\mu\lambda}^{(Q\bar{Q})}$ denote the $S$-wave, $P$-wave and $D$-wave multiplets of bottomonium, respectively. $H^{(Q\bar{q})}$ and $\bar{H}^{(\bar{Q}q)}$ are $(0^-,1^-)$ doublets of bottom and anti-bottom mesons respectively. $\mathcal{A^{\mu}}$ is the axial vector current of Nambu-Goldstone fields, which reads as $\mathcal{A}^{\mu}=(\xi^{\dagger}\partial^{\mu}\xi-\xi\partial^{\mu}\xi^{\dagger})/2$ with $\xi=e^{i\mathbb{P}/f_{\pi}}$. In addition, $\rho_{\mu}=ig_{V}\mathbb{V}_{\mu}/\sqrt{2}$ and $F_{\mu\nu}(\rho)=\partial_{\mu}\rho_{\nu}-\partial_{\nu}\rho_{\mu}+[\rho_{\mu},\rho_{\nu}]$ \cite{Huang:2017kkg,Wang:2020dya,Wang:2020bjt,Wang:2021hql}. For the detailed expressions of these symbols, we collect them into Appendix \ref{app01}.

After expanding the above Lagrangians, the concerned effective Lagrangians, as shown in Appendix \ref{app01}, can be obtained.
Thus, we can write down the concrete amplitudes corresponding to the diagrams in Figs. \ref{transitionUpseta}-\ref{transitionchibJomega}. For example,
with the Feynman rules listed in Appendix \ref{app02}, the amplitude of the first diagram in Fig. \ref{transitionUpseta} can be obtained as
\begin{equation}
\begin{split}
\mathcal{M}_{3D}^{(1)}=&i^3\int\frac{d^4q}{(2\pi)^4}g_{\Upsilon_1BB}\epsilon_{\Upsilon_{1}}^{\lambda}\epsilon_{\Upsilon}^{\nu*}
(q_{1\lambda}-q_{2\lambda})g_{\Upsilon BB^{*}}\varepsilon_{\mu\nu\alpha\beta}\\
&\times p_{2}^{\mu}(q_{2}^{\beta}-q^{\beta})g_{BB^{*}\eta^{(\prime)}}p_{1}^{\tau}
\frac{1}{q_{1}^{2}-m_{B}^{2}}\frac{1}{q_{2}^{2}-m_{\bar{B}}^{2}}\\
&\times\frac{-g_{\tau}^{\alpha}+q_{\tau}q^{\alpha}/m_{B^{*}}^2}{q^2-m_{B^{*}}^2}
\mathcal{F}^{2}(q^2,m_{B^{*}}^2).
\end{split}
\end{equation}
And then, the remaining amplitudes can be written down similarly.

Finally, the decay widths of the transitions of the $\Upsilon(10753)$ into a low-lying bottomonium by emitting a light pseudoscalar meson $\eta^{(\prime)}$ or $\omega$ can be evaluated by
\begin{equation}
\Gamma=\frac{1}{3}\frac{|\vec{p}|}{8\pi m^{2}}|\overline{\mathcal{M}^{\text{Total}}}|^{2},
\end{equation}
where the overbar denotes to sum over the polarizations of $\Upsilon(1S)$, $h_{b}(1P)$ or $\chi_{bJ}(1P)$ ($J$=0,1,2), and the vector meson $\omega$. The coefficient $1/3$ comes from averaging over polarizations of the initial state.
$\vec{p}$ is the momentum of the light meson.
The general expression of total amplitudes is
\begin{equation}
\mathcal{M}^{\text{Total}}=4\sum_{\text{i}=1}^{\text{i}_{max}}\mathcal{M}_{4S}^{\text{(i)}}\sin{\theta}
+4\sum_{\text{j}=1}^{\text{j}_{max}}\mathcal{M}_{3D}^{\text{(j)}}\cos{\theta}.
\end{equation}
Here, the superscript $\text{i(j)}$ denotes the $\text{i(j)}$-th amplitudes from bottomed meson loops in the above diagrams, while the subscripts $4S$ and $3D$ denote the contributions from $\Upsilon(4S)$ or $\Upsilon(3D)$ components, respectively, $\theta=33\degree$ is used in this work.
In addition, the factor 4 comes from the charge conjugation and the isospin transformations on the bridged $B^{(*)}$ meson.

\section{numerical results}
\label{sec4}
Before presenting the numerical results, we need to illustrate how to determine the relevant coupling constants.
The coupling constants $g_{\Upsilon_1B^{(*)}B^{(*)}}$ depicting the coupling between $\Upsilon(3D)$ and a pair of bottomed mesons are extracted from the corresponding decays widths, which are quoted from Ref. \cite{Wang:2018rjg}. The decay widths and the corresponding coupling constants are listed in Table \ref{coupling constants}. As for $g_{\Upsilon(4S)BB^{(*)}}$, the $g_{\Upsilon(4S)BB}$ is determined by the corresponding partial decay width given in Ref. \cite{Wang:2018rjg}, while the $g_{\Upsilon(4S)BB^{*}}$ can be fixed by $g_{\Upsilon(4S)BB}$ and Eq. (\ref{relationships between coupling constants}) in heavy quark symmetry. Numerically, $g_{\Upsilon(4S)BB}=13.224$ and $g_{\Upsilon(4S)BB^{*}}=1.251\ \text{GeV}^{-1}$ are used in this work.

\begin{table}[htbp]\centering
\caption{The decay widths of $\Upsilon(3D)\to B^{(*)}\bar{B}^{(*)}$ given in Ref. \cite{Wang:2018rjg} and the extracted coupling constants $g_{\Upsilon_1B^{(*)}B^{(*)}}$ of $\Upsilon(3D)$ coupling with $B^{(*)}\bar{B}^{(*)}$.}
\label{coupling constants}
\renewcommand\arraystretch{1.05}
\begin{tabular*}{86mm}{l@{\extracolsep{\fill}}ccc}
\toprule[1pt]
\toprule[0.5pt]
Channel                        &Decay width (MeV)   &Coupling constant\\
\midrule[0.5pt]
$B\bar{B}$                     &5.47                &3.480\\
$B\bar{B}^{*}\text{+c.c.}$     &15.2                &0.393 GeV$^{-1}$\\
$B^{*}\bar{B}^{*}$             &33.4                &4.210\\
\bottomrule[0.5pt]
\bottomrule[1pt]
\end{tabular*}
\end{table}

For $g_{\Upsilon B^{(*)}B^{(*)}}$, $g_{h_{b}B^{(*)}B^{(*)}}$ and $g_{\chi_{bJ}B^{(*)}B^{(*)}}$ defined in Eq. (\ref{SBB}), Eq. (\ref{PBB}) and Eq. (\ref{PJBB}), respectively, the symmetry implied by the heavy quark effective theory is needed to be considered. Under this symmetry, these coupling constants are related to each others through the global constants $g_{S}$ and $g_{P}$, which are expressed as
\begin{equation}
\begin{split}
&\frac{g_{\Upsilon BB}}{m_{B}}=
\frac{g_{\Upsilon BB^{*}}m_{\Upsilon}}{\sqrt{m_{B}m_{B^{*}}}}=
\frac{g_{\Upsilon B^{*}B^{*}}}{m_{B^{*}}}=
2g_{S}\sqrt{m_{\Upsilon}},\\
&\frac{g_{h_{b}BB^{*}}}{\sqrt{m_{B}m_{B^{*}}}}=
g_{h_{b}B^{*}B^{*}}\frac{m_{h_{b}}}{m_{B^{*}}}=
2g_{P}\sqrt{m_{h_{b}}},\\
&\frac{g_{\chi_{b0}BB}}{\sqrt{3}m_{B}}=\frac{\sqrt{3}g_{\chi_{b0}B^{*}B^{*}}}{m_{B^{*}}}=2\sqrt{m_{\chi_{b0}}}g_{P},\\
&g_{\chi_{b1}BB^{*}}=2\sqrt{2}\sqrt{m_{B}m_{B^{*}}m_{\chi_{b1}}}g_{P},\\
&g_{\chi_{b2}BB}m_{B}=g_{\chi_{b2}BB^{*}}\sqrt{m_{B}m_{B^{*}}}m_{\chi_{b2}}=\frac{g_{\chi_{b2}B^{*}B^{*}}}{4m_{B^{*}}}=\sqrt{m_{\chi_{b2}}}g_{P},\\
\end{split}
\label{relationships between coupling constants}
\end{equation}
where $g_{S}=0.407$ GeV$^{-3/2}$ \cite{Huang:2018pmk}, and $g_{P}=-\sqrt{\frac{m_{\chi_{b0}}}{3}}\frac{1}{f_{\chi_{b0}}}$ with $f_{\chi_{b0}}=175\pm55$ MeV is obtained by QCD sum rule \cite{Veliev:2010gb}.

Additionally, $g_{B^{(*)}B^{(*)}\eta}$ can be expressed by the coupling constant $g_{H}$, i.e.,
\begin{equation}
\begin{split}
\frac{g_{BB^{*}\eta}}{\sqrt{m_{B}m_{B^{*}}}}=g_{B^{*}B^{*}\eta}=\frac{2g_{H}}{f_{\pi}}\alpha,\\
\frac{g_{BB^{*}\eta^{\prime}}}{\sqrt{m_{B}m_{B^{*}}}}=g_{B^{*}B^{*}\eta^{\prime}}=\frac{2g_{H}}{f_{\pi}}\beta,
\end{split}
\end{equation}
where $g_{H}=0.569$, $f_{\pi}=$ 131 MeV \cite{Wang:2016qmz,Huang:2018cco,Huang:2018pmk}, $\alpha$ and $\beta$ were defined in Eq. (\ref{alphabetagammadelta}). And $g_{B^{(*)}B^{(*)}\omega}$ is related to  the global coupling constant $g_{V}$, i.e.,
\begin{equation}
\begin{split}
g_{BB\omega}=g_{B^{*}B^{*}\omega}=\frac{\beta g_{V}}{2},\\
f_{BB^{*}\omega}=\frac{f_{B^{*}B^{*}\omega}}{m_{B^{*}}}=\frac{\lambda g_{V}}{2},
\end{split}
\end{equation}
where $\beta=0.9$, $\lambda=0.56$ GeV$^{-1}$, and $g_{V}=m_{\rho}/f_{\pi}$ \cite{Huang:2017kkg}.

\subsection{The result of processes with $\eta^{(\prime)}$ emission}

\begin{figure}[htbp]\centering
  \begin{tabular}{cc}
  \includegraphics[width=42mm]{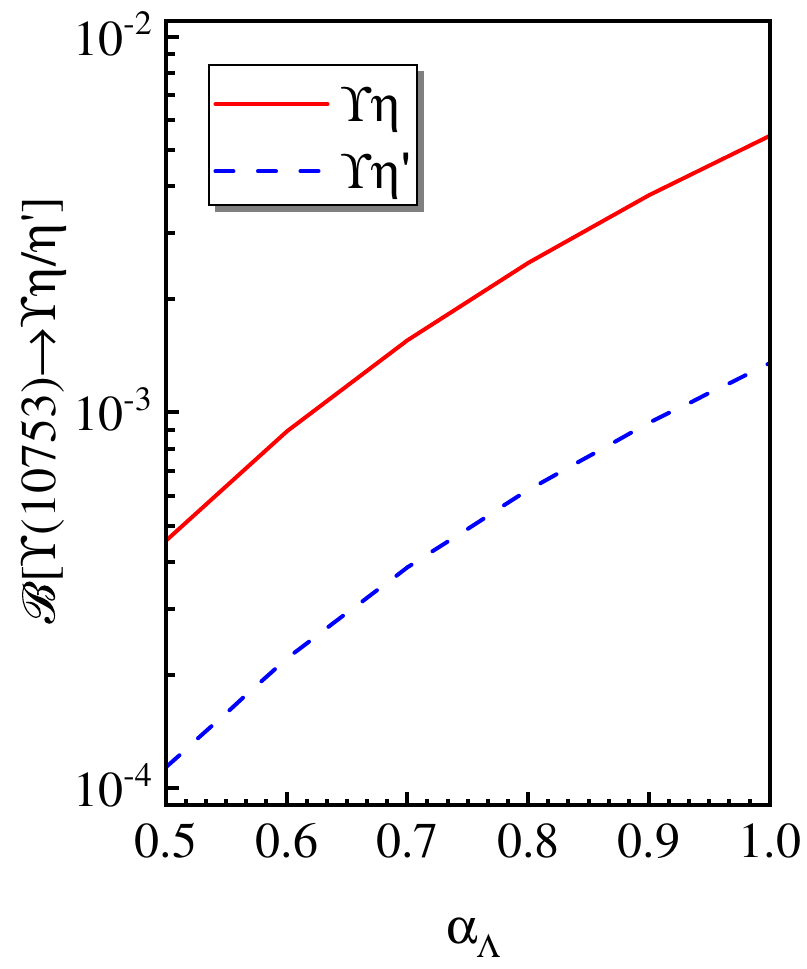}
  \includegraphics[width=42mm]{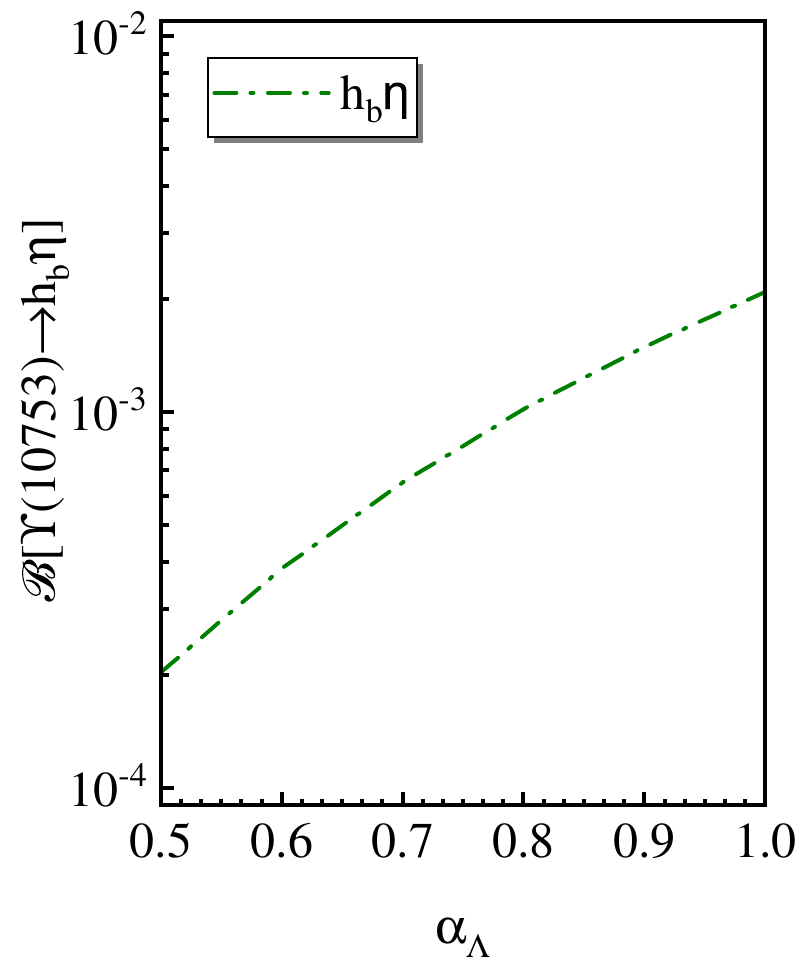}\\
  \end{tabular}
  \caption{The $\alpha_{\Lambda}$ dependence of the branching fractions $\mathcal{B}[\Upsilon(10753)\to\Upsilon(1S)\eta^{(\prime)}]$ (left panel) and $\mathcal{B}[\Upsilon(10753)\to h_{b}(1P)\eta]$ (right panel).}
\label{alphaeta}
\end{figure}

In this subsection, we present our results of the $\Upsilon(10753)\to\Upsilon(1S)\eta^{(\prime)}$, $\Upsilon(10753)\to h_{b}(1P)\eta$ processes.
Apart from the fixed coupling constants, there still exists a free parameter $\alpha_{\Lambda}$, which comes from the introduced form factor.
Following Ref. \cite{Wang:2016qmz,Huang:2017kkg,Huang:2018cco,Zhang:2018eeo}, in this work we still set $0.5\leq\alpha_{\Lambda}\leq1.0$ to calculate the decay widths and the corresponding branching fractions for the $\Upsilon(10753)\to\Upsilon(1S)\eta^{(\prime)}$ and $\Upsilon(10753)\to h_{b}(1P)\eta$ processes, and the numerical results are shown in Fig. \ref{alphaeta}.

Thus, the decay widths of $\Upsilon(10753)\to\Upsilon(1S)\eta^{(\prime)}$, $\Upsilon(10753)\to h_{b}(1P)\eta$ processes are
\begin{eqnarray*}
\Gamma[\Upsilon(10753)\to\Upsilon(1S)\eta]&=&(16.2 - 193.7)\ \text{keV},\\
\Gamma[\Upsilon(10753)\to\Upsilon(1S)\eta^{\prime}]&=&(4.04 - 48.0)\ \text{keV},\\
\Gamma[\Upsilon(10753)\to h_{b}(1P)\eta]&=&(7.23 - 74.2)\ \text{keV},
\end{eqnarray*}
which are transferred from these branching fractions
\begin{eqnarray*}
\mathcal{B}[\Upsilon(10753)\to\Upsilon(1S)\eta]&=&(0.46 - 5.46) \times 10^{-3},\\
\mathcal{B}[\Upsilon(10753)\to\Upsilon(1S)\eta^{\prime}]&=&(0.11 - 1.35) \times 10^{-3},\\
\mathcal{B}[\Upsilon(10753)\to h_{b}(1P)\eta]&=&(0.20 - 2.09) \times 10^{-3}.
\end{eqnarray*}

Our results show that the branching fractions of the $\Upsilon(10753)\to\Upsilon(1S)\eta$, $\Upsilon(10753)\to h_{b}(1P)\eta$ processes can reach up to $10^{-3}$ and the branching ratio of $\Upsilon(10753)\to\Upsilon(1S)\eta^{\prime}$ is around $10^{-4}$, which indicates that these discussed transitions can be accessible at the Belle II experiment.
In addition, we provide the ratio $R_{\eta^{\prime}/\eta}=\mathcal{B}[\Upsilon(10753)\to\Upsilon(1S)\eta^{\prime}]/\mathcal{B}[\Upsilon(10753)\to\Upsilon(1S)\eta]$ as
\begin{eqnarray*}
R_{\eta^{\prime}/\eta}\approx0.25 ,
\end{eqnarray*}
which is weakly dependent on the cutoff parameter $\alpha_{\Lambda}$.
The behavior of $R_{\eta^{\prime}/\eta}$ can be further tested in future experiments.

\subsection{The result of the processes with $\omega$ emission}

\begin{figure}[htbp]\centering
  \begin{tabular}{cc}
  \includegraphics[width=42mm]{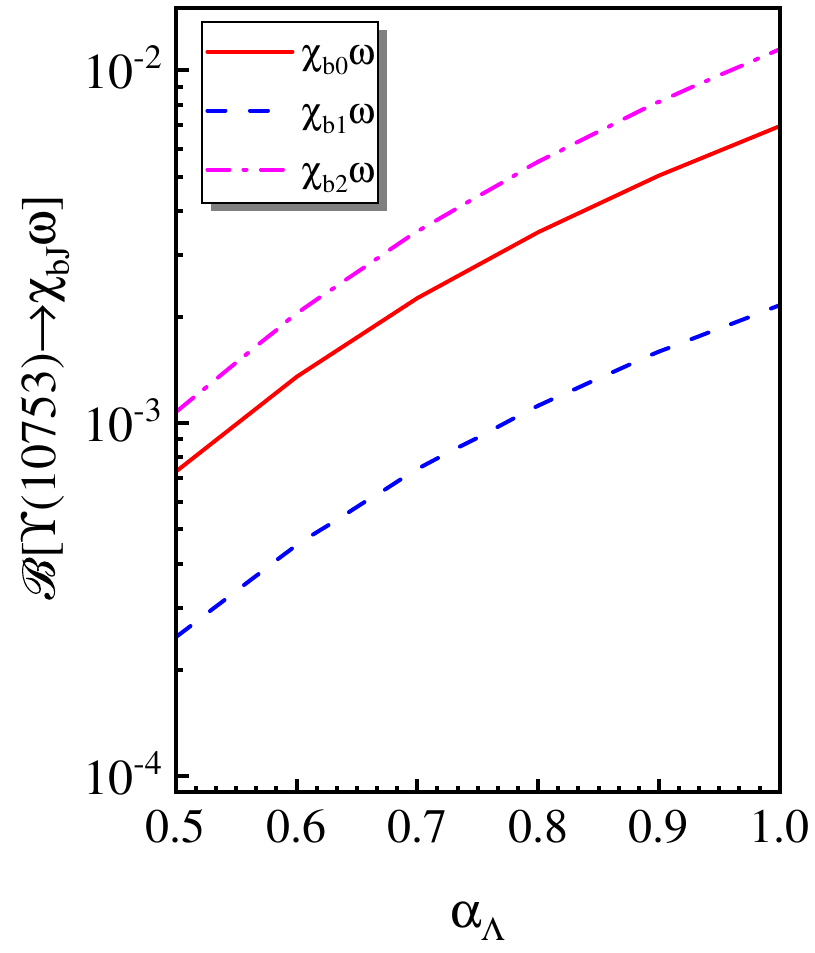}
  \includegraphics[width=42mm]{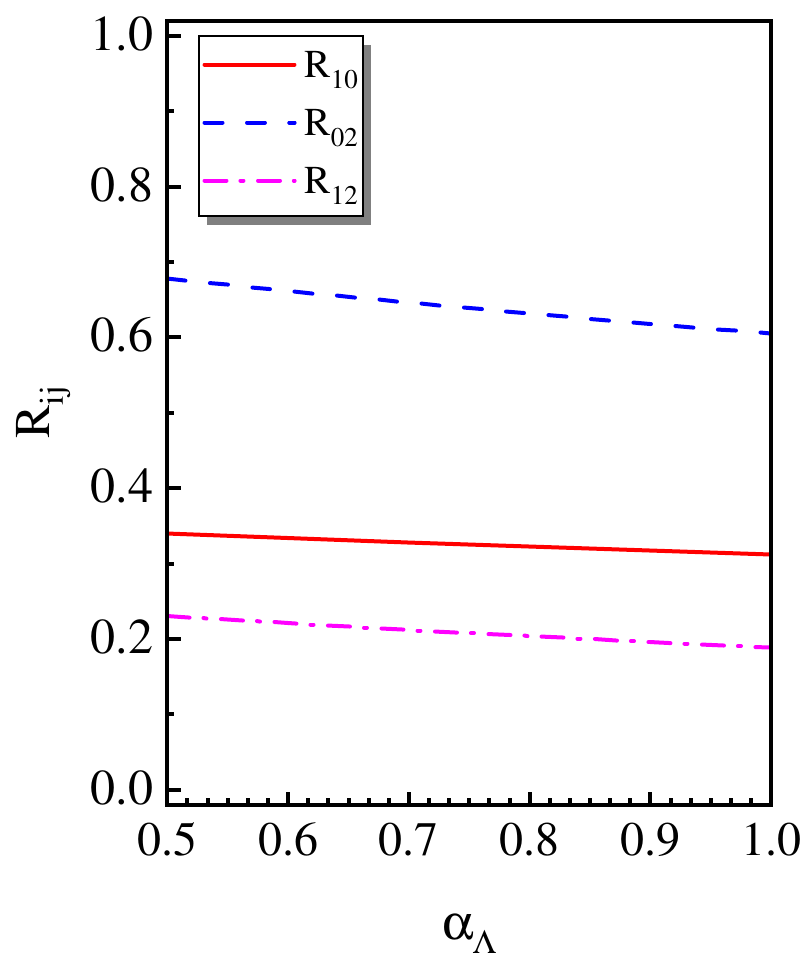}\\
  \end{tabular}
  \caption{The $\alpha_{\Lambda}$ dependence of the branching fractions $\mathcal{B}[\Upsilon(10753)\to\chi_{bJ}\omega]$ (left panel) and the $\alpha_{\Lambda}$ dependence of the ratios $R_{ij}=\mathcal{B}[\Upsilon(10753)\to\chi_{bi}\omega]/\mathcal{B}[\Upsilon(10753)\to\chi_{bj}\omega]$ (right panel).}
\label{alphaomega}
\end{figure}

Next, we investigate the hadronic loop contribution to the $\Upsilon(10753)\to\chi_{bJ}\omega$ ($J=0,1,2$) decays, where the $\alpha_{\Lambda}$ dependence of branching fractions are shown in Fig. \ref{alphaomega}.

From Fig. \ref{alphaomega} we can get that the corresponding branching fractions are
\begin{eqnarray*}
\mathcal{B}[\Upsilon(10753)\to\chi_{b0}\omega]&=&(0.73 - 6.94)\times 10^{-3},\\
\mathcal{B}[\Upsilon(10753)\to\chi_{b1}\omega]&=&(0.25 - 2.16)\times 10^{-3},\\
\mathcal{B}[\Upsilon(10753)\to\chi_{b2}\omega]&=&(1.08 - 11.5)\times 10^{-3}.
\end{eqnarray*}
Thus the corresponding decay widths are
\begin{eqnarray*}
\Gamma[\Upsilon(10753)\to\chi_{b0}\omega]&=&(25.9 - 246.5)\ \text{keV},\\
\Gamma[\Upsilon(10753)\to\chi_{b1}\omega]&=&(8.79 - 76.8)\ \text{keV},\\
\Gamma[\Upsilon(10753)\to\chi_{b2}\omega]&=&(38.2 - 407.7)\ \text{keV}.
\end{eqnarray*}

Additionally, we also notice that the behavior of the ratios $R_{ij}=\mathcal{B}[\Upsilon(10753)\to\chi_{bi}\omega]/\mathcal{B}[\Upsilon(10753)\to\chi_{bj}\omega]$ (see Fig. \ref{alphaomega}) are also weakly dependent on the parameter $\alpha_{\Lambda}$, i.e.,
\begin{eqnarray*}
R_{10}&=&\frac{\mathcal{B}[\Upsilon(10753)\to\chi_{b1}\omega]}{\mathcal{B}[\Upsilon(10753)\to\chi_{b0}\omega]} = (0.33 - 0.35),\\
R_{12}&=&\frac{\mathcal{B}[\Upsilon(10753)\to\chi_{b1}\omega]}{\mathcal{B}[\Upsilon(10753)\to\chi_{b2}\omega]} = (0.18 - 0.22),\\
R_{02}&=&\frac{\mathcal{B}[\Upsilon(10753)\to\chi_{b0}\omega]}{\mathcal{B}[\Upsilon(10753)\to\chi_{b2}\omega]} = (0.55 - 0.63).
\end{eqnarray*}
Obviously, experimental measurement for them will be an interesting issue for future experiments such as Belle II.

\section{Summary}
\label{sec5}
Recently, the Belle Collaboration released a new analysis on the $e^{+}e^{-}\to\Upsilon(nS)\pi^{+}\pi^{-}$ processes ($n=1,2,3$) and found the evidence of a new structure near 10.75 GeV \cite{Abdesselam:2019gth}, whose mass and width are fitted as $M=(10752.7\pm5.9^{+0.7}_{-1.1})$ MeV and $\Gamma=(35.5^{+17.6 +3.9}_{-11.3 -3.3})$ MeV, respectively. Although the width of this state is consistent with the prediction of $\Upsilon(3D)$ given by Ref. \cite{Wang:2018rjg}, its mass can not match any of the previous theoretical results of $\Upsilon(3D)$ \cite{Segovia:2016xqb,Wang:2018rjg,Godfrey:2015dia,Badalian:2008ik,Badalian:2009bu}.

To solve the mass problem of this $\Upsilon(10753)$, based on the experience on the charmonium family \cite{Rosner:2001nm,Wang:2019mhs,Wang:2020prx,Qian:2021gby}, $4S-3D$ mixing is introduced. It is found that when the mixing angle is around $(33 \pm 4)\degree$, the masses of the $\Upsilon(10580)$ and $\Upsilon(10753)$ can be reproduced simultaneously. Besides, the obtained masses of pure $4S$ and $3D$ states are consistent with \cite{Segovia:2016xqb,Wang:2018rjg,Godfrey:2015dia,Badalian:2008ik,Badalian:2009bu}. Thus, in our view, the newly observed $\Upsilon(10753)$ can be assigned as a bottomonium state with $4S$-$3D$ mixing.

To further understand the nature of this $\Upsilon(10753)$, study on its decay behavior is necessary. Since the study of its hidden-bottom hadronic decays is still absent, in this work we carry out a study on some hidden-bottom decays of the $\Upsilon(10753)$, which include $\Upsilon(10753)\to\Upsilon(1S)\eta^{(\prime)}$, $\Upsilon(10753)\to h_{b}(1P)\eta$ and $\Upsilon(10753)\to\chi_{bJ}\omega$ ($J$=0,1,2).

Due to the typical feature that the $\Upsilon(10753)$ has the mass above the $B\bar{B}$ threshold, the coupled-channel effect should be considered, thus in this work we absorb the hadronic loop mechanism into the realistic study on the $\Upsilon(10753)$. We obtain the partial decay widths and the corresponding branching fractions of these discussed hidden-bottom decays, which show that $\Upsilon(10753)\to\Upsilon(1S)\eta^{(\prime)}$, $\Upsilon(10753)\to h_{b}(1P)\eta$ and $\Upsilon(10753)\to\chi_{bJ}\omega$ ($J$=0,1,2) have considerable decay rates, which we think are accessible in future experiments such us Belle II. Furthermore, we obtain some typical values which are almost independent the free parameter introduced by the form factor, which can be important values for us to study the decay behavior of this $\Upsilon(10753)$ and understand its nature.

In conclusion, it is obvious that future experimental search for these discussed hidden-bottom hadronic decays of the $\Upsilon(10753)$ will be an intriguing task, by which the nature of the $\Upsilon(10753)$ should be decoded. Thus, We strongly encourage the running Belle II experiment to pay attention to this issue.

\section*{ACKNOWLEDGMENTS}
This work is supported by the China National Funds for Distinguished Young Scientists under Grant No. 11825503, National Key Research and Development Program of China under Contract No. 2020YFA0406400, and the 111 Project under Grant No. B20063, the National Natural Science Foundation of China under Grant No. 12047501.

\appendix

\section{}
\label{app01}

In this appendix, we give the adopted Lagrangians in detail.
The $S^{(Q\bar{Q})}$, $P^{(Q\bar{Q})\mu}$ and $D_{\mu\lambda}^{(Q\bar{Q})}$ denote the $S$-wave, $P$-wave and $D$-wave multiplets of the bottomonia with expressions
\begin{equation}
S^{(Q\bar{Q})}=\frac{1+\slashed{\upsilon}}{2}\left[\Upsilon^{\mu}\gamma_{\mu}-\eta_{b}\gamma_{5}\right]\frac{1-\slashed{\upsilon}}{2},
\end{equation}
\begin{equation}
\begin{split}
P^{(Q\bar{Q})\mu}=&\frac{1+\slashed{\upsilon}}{2}\left[\chi_{b2}^{\mu\alpha}\gamma_{\alpha}+\frac{1}{\sqrt{2}}\varepsilon^{\mu\alpha\beta\gamma}\upsilon_{\alpha}\gamma_{\beta}\chi_{b1\gamma}\right.\\
&\left.+\frac{1}{\sqrt{3}}\left(\gamma^{\mu}-\upsilon^{\mu}\right)\chi_{b0}+h_{b}^{\mu}\gamma_{5}\right]\frac{1-\slashed{\upsilon}}{2},
\end{split}
\end{equation}
\begin{equation}
\begin{split}
D^{(Q\bar{Q})\mu\nu}=&\frac{1+\slashed{\upsilon}}{2}\left[\Upsilon_{3}^{\mu\nu\alpha}\gamma_{\alpha}+\frac{1}{\sqrt{6}}\left(\varepsilon^{\mu\alpha\beta\rho}\upsilon_{\alpha}\gamma_{\beta}\Upsilon_{2\rho}^{\nu}+\varepsilon^{\nu\alpha\beta\rho}\upsilon_{\alpha}\gamma_{\beta}\right.\right.\\
&\left.\left.\times\Upsilon_{2\rho}^{\mu}\right)+\frac{\sqrt{15}}{10}\left[\left(\gamma^{\mu}-\upsilon^{\mu}\right)\Upsilon_{1}^{\nu}+\left(\gamma^{\nu}-\upsilon^{\nu}\right)\Upsilon_{1}^{\mu}\right]\right.\\
&\left.-\frac{1}{\sqrt{15}}\left(g^{\mu\nu}-\upsilon^{\mu}\upsilon^{\nu}\right)\gamma_{\alpha}\Upsilon_{1}^{\alpha}+\eta_{b2}^{\mu\nu}\gamma_{5}\right]\frac{1-\slashed{\upsilon}}{2},
\end{split}
\end{equation}
respectively. $H^{(Q\bar{q})}$ represents the spin doublet of bottom meson filed $(\mathcal{B}, \mathcal{B}^{*})$ and $H^{(\bar{Q}q)}$ corresponds to the anti-meson counterpart. Their concrete expressions are \cite{Xu:2016kbn,Duan:2021bna}
\begin{equation}
\begin{split}
H^{(Q\bar{q})}=&\frac{1+\slashed{\upsilon}}{2}(\mathcal{B}^{*\mu}\gamma_{\mu}+i\mathcal{B}\gamma_{5}),\\
H^{(\bar{Q}q)}=&(\bar{\mathcal{B}}^{*\mu}\gamma_{\mu}+i\bar{\mathcal{B}}\gamma_{5})\frac{1-\slashed{\upsilon}}{2}
\end{split}
\end{equation}
with $\bar{H}^{(Q\bar{q})}=\gamma_{0}H^{\dagger(Q\bar{q})}\gamma_{0}$ and $\bar{H}^{(\bar{Q}q)}=\gamma_{0}H^{\dagger(\bar{Q}q)}\gamma_{0}$.

In addition, the pseudoscalar octet $\mathbb{P}$ reads as
\begin{equation}
\begin{split}
\mathbb{P}&=
\begin{pmatrix}
\frac{\pi^{0}}{\sqrt{2}}+\alpha\eta+\beta\eta^{\prime}&\pi^{+}&K^{+}\\
\pi^{-}&-\frac{\pi^{0}}{\sqrt{2}}+\alpha\eta+\beta\eta^{\prime}&K^{0}\\
K^{-}&\bar{K}^{0}&\gamma\eta+\delta\eta^{\prime}
\end{pmatrix}\\
\end{split}
\end{equation}
with
\begin{equation}
\begin{split}
&\alpha=\frac{\cos\theta-\sqrt{2}\sin\theta}{\sqrt{6}},\quad \beta=\frac{\sin\theta+\sqrt{2}\cos\theta}{\sqrt{6}},\\
&\gamma=\frac{-2\cos\theta-\sqrt{2}\sin\theta}{\sqrt{6}},\quad \delta=\frac{-2\sin\theta+\sqrt{2}\cos\theta}{\sqrt{6}},
\label{alphabetagammadelta}
\end{split}
\end{equation}
and the mixing angle $\theta=-19.1\degree$ \cite{Wang:2016qmz, Huang:2018pmk, Coffman:1988ve, Jousset:1988ni},
while the expression of $\mathbb{V}$ reads as
\begin{equation}
\begin{split}
\mathbb{V}&=
\begin{pmatrix}
\frac{1}{\sqrt{2}}(\rho^{0}+\omega)&\rho^{+}&K^{*+}\\
\rho^{-}&\frac{1}{\sqrt{2}}(-\rho^{0}+\omega)&K^{*0}\\
K^{*-}&\bar{K}^{*0}&\phi
\end{pmatrix}.\\
\end{split}
\end{equation}

With the above preparation, the compact Lagrangians in Eq. (\ref{totalLagrangians}) can be expanded as
\begin{equation}
\begin{split}
\mathcal{L}_{\Upsilon\mathcal{B}^{(*)}\mathcal{B}^{(*)}}=&\
ig_{\Upsilon\mathcal{B}\mathcal{B}}\Upsilon^{\mu}(\partial_{\mu}\mathcal{B}^{\dagger}\mathcal{B}-\mathcal{B}^{\dagger}\partial_{\mu}\mathcal{B})\\
&+g_{\Upsilon\mathcal{B}\mathcal{B}^{*}}\varepsilon_{\mu\nu\alpha\beta}\partial^{\mu}\Upsilon^{\nu}(\mathcal{B}^{*\alpha}\overset\leftrightarrow\partial^{\beta}\mathcal{B}^{\dagger}-\mathcal{B}\overset\leftrightarrow\partial^{\beta}\mathcal{B}^{*\alpha\dagger})\\
&+ig_{\Upsilon\mathcal{B}^{*}\mathcal{B}^{*}}\Upsilon^{\mu}(\mathcal{B}^{*\nu}\partial_{\nu}\mathcal{B}_{\mu}^{*\dagger}-\mathcal{B}^{*\dagger\nu}\partial_{\nu}\mathcal{B}_{\mu}^{*}-\mathcal{B}^{*\nu}\overset\leftrightarrow\partial_{\mu}\mathcal{B}_{\nu}^{*\dagger}),
\label{SBB}
\end{split}
\end{equation}
\begin{equation}
\begin{split}
\mathcal{L}_{h_{b}\mathcal{B}^{(*)}\mathcal{B}^{(*)}}=&
-g_{h_{b}\mathcal{B}\mathcal{B}^{*}}(\mathcal{B}^{\dagger}\mathcal{B}_{\mu}^{*}+\mathcal{B}_{\mu}^{*\dagger}\mathcal{B})h_{b}^{\mu}\\
&-ig_{h_{b}\mathcal{B}^{*}\mathcal{B}^{*}}\varepsilon_{\alpha\beta\mu\nu}\mathcal{B}^{*\alpha}\mathcal{B}^{*\dagger\beta}\partial^{\nu}h_{b}^{\mu},
\label{PBB}
\end{split}
\end{equation}
\begin{equation}
\begin{split}
\mathcal{L}_{\chi_{bJ}\mathcal{B}^{(*)}\mathcal{B}^{(*)}}=&\mathrm{i}g_{\chi_{b0}\mathcal{B}\mathcal{B}}\chi_{b0}\mathcal{B}\mathcal{B}^{\dagger}
-ig_{\chi_{b0}\mathcal{B}^{*}\mathcal{B}^{*}}\chi_{b0}\mathcal{B}^{*\mu}\mathcal{B}^{*\dagger}_{\mu}\\
&-ig_{\chi_{b1}\mathcal{B}\mathcal{B}^{*}}\chi_{b1}^{\mu}(\mathcal{B}^{*}_{\mu}\mathcal{B}^{\dagger}-\mathcal{B}\mathcal{B}^{*\dagger}_{\mu})\\
&+ig_{\chi_{b2}\mathcal{B}\mathcal{B}}\chi_{b2}^{\mu\nu}\partial_{\mu}\mathcal{B}\partial_{\nu}\mathcal{B}^{\dagger}
+ig_{\chi_{b2}\mathcal{B}^{*}\mathcal{B}^{*}}\chi_{b2}^{\mu\nu}\mathcal{B}_{\mu}^{*}\mathcal{B}_{\nu}^{*\dagger}\\
&+g_{\chi_{b2}\mathcal{B}\mathcal{B}^{*}}\varepsilon_{\mu\nu\alpha\beta}\partial^{\alpha}\chi_{b2}^{\mu\rho}(\partial_{\rho}\mathcal{B}^{*\nu}\partial^{\beta}\mathcal{B}^{\dagger}-\partial^{\beta}\mathcal{B}\partial_{\rho}\mathcal{B}^{*\dagger\nu}),
\label{PJBB}
\end{split}
\end{equation}
\begin{equation}
\begin{split}
\mathcal{L}_{\Upsilon_{1}\mathcal{B}^{(*)}\mathcal{B}^{(*)}}=&\
ig_{\Upsilon_{1}\mathcal{B}\mathcal{B}}\Upsilon_{1}^{\mu}(\partial_{\mu}\mathcal{B}^{\dagger}\mathcal{B}-\mathcal{B}^{\dagger}\partial_{\mu}\mathcal{B})\\
&-g_{\Upsilon_{1}\mathcal{B}\mathcal{B}^{*}}\varepsilon_{\mu\nu\alpha\beta}\partial^{\nu}\Upsilon_{1}^{\alpha}(\mathcal{B}^{*\beta}\overset\leftrightarrow\partial^{\mu}\mathcal{B}^{\dagger}-\mathcal{B}\overset\leftrightarrow\partial^{\mu}\mathcal{B}^{*\beta\dagger})\\
&+ig_{\Upsilon_{1}\mathcal{B}^{*}\mathcal{B}^{*}}\Upsilon_{1}^{\mu}(-4\mathcal{B}^{*\nu}\overset\leftrightarrow\partial_{\mu}\mathcal{B}_{\nu}^{*\dagger}+\mathcal{B}^{*\nu}\partial_{\nu}\mathcal{B}_{\mu}^{*\dagger}\\
&-\mathcal{B}^{*\nu\dagger}\partial_{\nu}\mathcal{B}_{\mu}^{*}),
\label{DBB}
\end{split}
\end{equation}
\begin{equation}
\begin{split}
\mathcal{L}_{\mathcal{B}^{(*)}\mathcal{B}^{(*)}\mathbb{P}}=&\ ig_{\mathcal{B}\mathcal{B}^{*}\mathbb{P}}(\mathcal{B}_{\mu}^{*\dagger}\mathcal{B}-\mathcal{B}^{\dagger}\mathcal{B}_{\mu}^{*})\partial^{\mu}\mathbb{P}\\
&-g_{\mathcal{B}^{*}\mathcal{B}^{*}\mathbb{P}}\varepsilon_{\mu\nu\alpha\beta}\partial^{\mu}\mathcal{B}^{*\dagger\nu}\partial^{\alpha}\mathcal{B}^{*\beta}\mathbb{P},
\label{BBP}
\end{split}
\end{equation}
\begin{equation}
\begin{split}
\mathcal{L}_{\mathcal{B}^{(*)}\mathcal{B}^{(*)}\mathbb{V}}=&
-ig_{\mathcal{B}\mathcal{B}\mathbb{V}}\mathcal{B}^{\dagger}_{\text{i}}\overset\leftrightarrow{\partial}^{\mu}\mathcal{B}^{\text{j}}(\mathbb{V}_{\mu})_{\text{j}}^{\text{i}}\\
&-2f_{\mathcal{B}\mathcal{B}^{*}\mathbb{V}}\varepsilon_{\mu\nu\alpha\beta}\partial^{\mu}(\mathbb{V}^{\nu})_{\text{j}}^{\text{i}}(\mathcal{B}^{\dagger}_{\text{i}}\overset\leftrightarrow{\partial}^{\alpha}
\mathcal{B}^{*\beta\text{j}}-\mathcal{B}^{*\beta\dagger}_{\text{i}}\overset\leftrightarrow{\partial}^{\alpha}\mathcal{B}^{\text{j}})\\
&+ig_{\mathcal{B}^{*}\mathcal{B}^{*}\mathbb{V}}\mathcal{B}_{\text{i}}^{*\nu\dagger}\overset\leftrightarrow{\partial}^{\mu}\mathcal{B}_{\nu}^{*\text{j}}(\mathbb{V}_{\mu})^{\text{i}}_{\text{j}}\\
&+4if_{\mathcal{B}^{*}\mathcal{B}^{*}\mathbb{V}}\mathcal{B}_{\text{i}}^{*\mu\dagger}(\partial_{\mu}\mathbb{V}_{\nu}-\partial_{\nu}\mathbb{V}_{\mu})^{\text{i}}_{\text{j}}
\mathcal{B}^{*\nu j},
\label{BBV}
\end{split}
\end{equation}
where $\mathcal{B}^{(*)\dagger}$ and $\mathcal{B}^{(*)}$ are defined as $\mathcal{B}^{(*)\dagger} = (B^{(*)+}, B^{(*)0}, B_{s}^{(*)0})$ and $\mathcal{B}^{(*)} = (B^{(*)-}, \bar{B}^{(*)0}, \bar{B}_{s}^{(*)0})^{T}$, respectively.

\section{}
\label{app02}
In this appendix, the Feynman rules for each interaction vertex involved in our calculation are collected.
We list the concrete information as follows
\begin{eqnarray}
\raisebox{-15pt}{\includegraphics[width=0.15%
\textwidth]{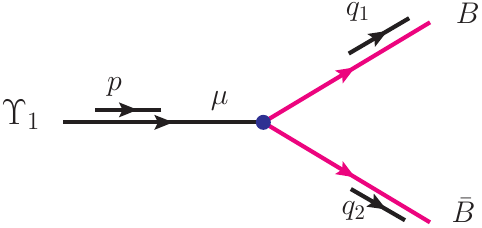}}           &\widehat{=}
&-g_{\Upsilon_1BB}\epsilon_{\Upsilon_1}^{\mu}(q_{1\mu}-q_{2\mu}),\\
\raisebox{-15pt}{\includegraphics[width=0.15%
\textwidth]{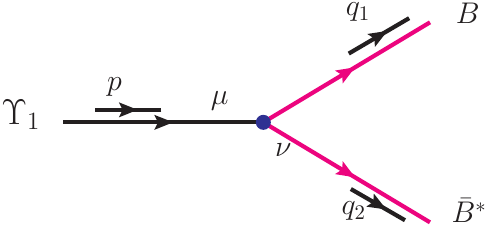}}       &\widehat{=}
&g_{\Upsilon_1BB^*}\varepsilon_{\alpha\beta\mu\nu}\epsilon_{\Upsilon_1}^{\mu}p^{\beta}(q_2^{\alpha}-q_1^{\alpha})\epsilon_{\bar{B}^{*}}^{*\nu},\\
\raisebox{-15pt}{\includegraphics[width=0.15%
\textwidth]{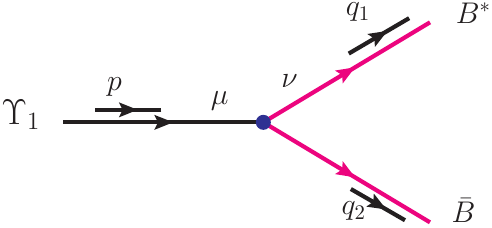}}       &\widehat{=}
&g_{\Upsilon_1BB^*}\varepsilon_{\alpha\beta\mu\nu}\epsilon_{\Upsilon_1}^{\mu}p^{\beta}(q_1^{\alpha}-q_2^{\alpha})\epsilon_{B^{*}}^{*\nu},\\
\raisebox{-15pt}{\includegraphics[width=0.15%
\textwidth]{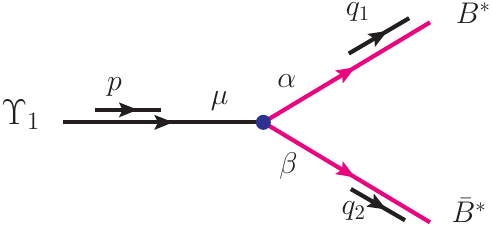}}   &\widehat{=}
&\begin{split}
&-g_{\Upsilon_1B^*B^*}\epsilon_{\Upsilon_1}^{\mu}(-4g_{\alpha\beta}(q_{1\mu}-q_{2\mu})\\
&+g_{\alpha\mu}q_{1\beta}-g_{\beta\mu}q_{2\alpha})\epsilon_{B^{*}}^{*\alpha}\epsilon_{\bar{B}^{*}}^{*\beta},
\end{split}\\
\raisebox{-15pt}{\includegraphics[width=0.15%
\textwidth]{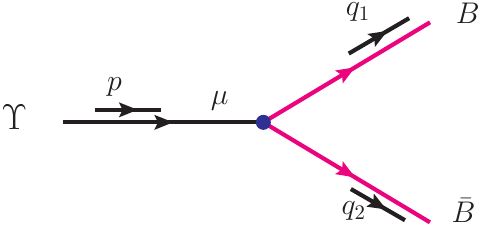}}           &\widehat{=}
&-g_{\Upsilon BB}\epsilon_{\Upsilon}^{\mu}(q_{1\mu}-q_{2\mu}),\\
\raisebox{-15pt}{\includegraphics[width=0.15%
\textwidth]{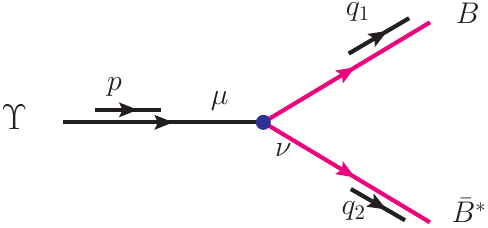}}       &\widehat{=}
&g_{\Upsilon BB^*}\varepsilon_{\alpha\beta\mu\nu}\epsilon_{\Upsilon}^{\mu}p^{\beta}(q_2^{\alpha}-q_1^{\alpha})\epsilon_{\bar{B}^{*}}^{*\nu},\\
\raisebox{-15pt}{\includegraphics[width=0.15%
\textwidth]{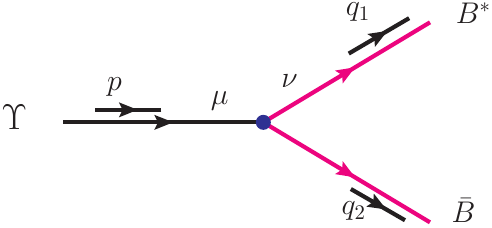}}       &\widehat{=}
&g_{\Upsilon BB^*}\varepsilon_{\alpha\beta\mu\nu}\epsilon_{\Upsilon}^{\mu}p^{\beta}(q_1^{\alpha}-q_2^{\alpha})\epsilon_{B^{*}}^{*\nu},\\
\raisebox{-15pt}{\includegraphics[width=0.15%
\textwidth]{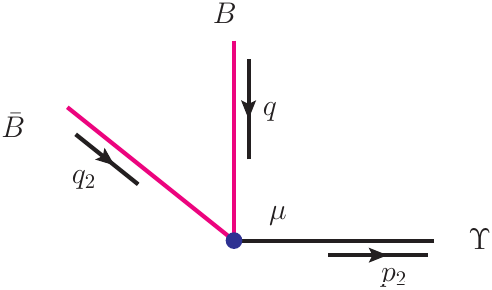}}           &\widehat{=}
&g_{\Upsilon BB}\epsilon_{\Upsilon}^{*\mu}(q_{2\mu}-q_{\mu}),\\
\raisebox{-15pt}{\includegraphics[width=0.15%
\textwidth]{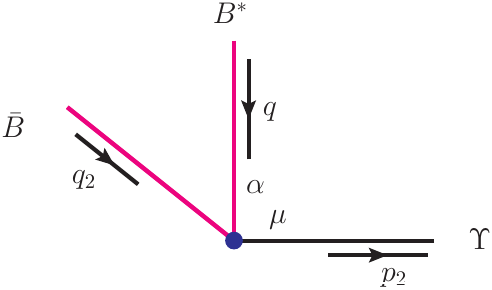}}       &\widehat{=}
&g_{\Upsilon BB^*}\varepsilon_{\mu\nu\alpha\beta}\epsilon_{\Upsilon}^{*\mu}p_{2}^{\nu}(q^{\beta}-q_2^{\beta})\epsilon_{B^{*}}^{\alpha},\\
\raisebox{-15pt}{\includegraphics[width=0.15%
\textwidth]{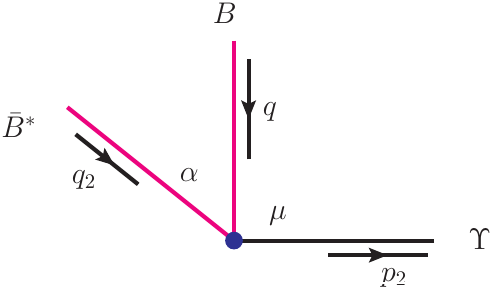}}       &\widehat{=}
&g_{\Upsilon BB^*}\varepsilon_{\mu\nu\alpha\beta}\epsilon_{\Upsilon}^{*\mu}p_{2}^{\nu}(q_2^{\beta}-q^{\beta})\epsilon_{\bar{B}^{*}}^{\alpha},\\
\raisebox{-15pt}{\includegraphics[width=0.15%
\textwidth]{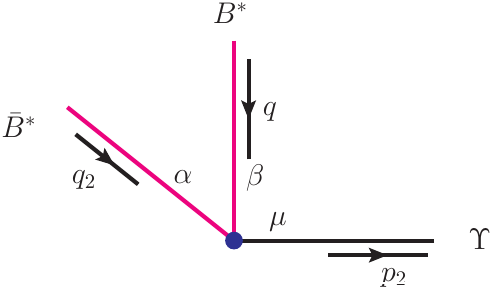}}   &\widehat{=}
&\begin{split}
&g_{\Upsilon B^*B^*}\epsilon_{\Upsilon}^{*\mu}(g_{\alpha\mu}q_{2\beta}-g_{\beta\mu}q_{\alpha}\\
&-g_{\alpha\beta}q_{2\mu}+g_{\alpha\beta}q_{\mu})\epsilon_{\bar{B}^{*}}^{\alpha}\epsilon_{B^{*}}^{\beta},
\end{split}\\
\raisebox{-15pt}{\includegraphics[width=0.15%
\textwidth]{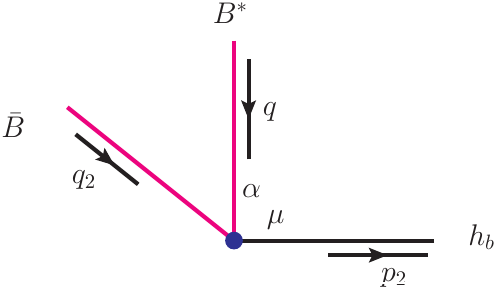}}           &\widehat{=}
&-g_{h_bBB^*}\epsilon_{h_b}^{*\mu}\epsilon_{B^{*}}^{\alpha}g_{\mu\alpha},\\
\raisebox{-15pt}{\includegraphics[width=0.15%
\textwidth]{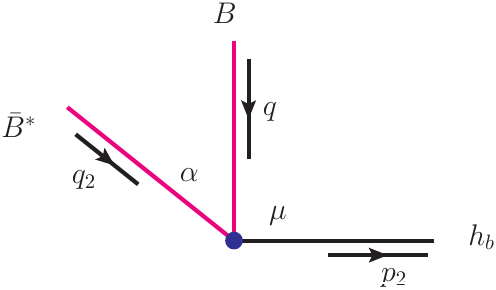}}           &\widehat{=}
&-g_{h_bBB^*}\epsilon_{h_b}^{*\mu}\epsilon_{\bar{B}^{*}}^{\alpha}g_{\mu\alpha},\\
\raisebox{-15pt}{\includegraphics[width=0.15%
\textwidth]{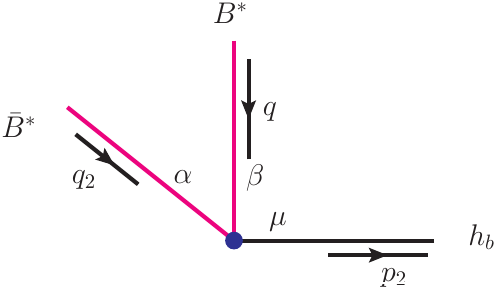}}       &\widehat{=}
&g_{h_bB^*B^*}\varepsilon_{\beta\alpha\mu\nu}\epsilon_{h_b}^{*\mu}p_{2}^{\nu}\epsilon_{\bar{B}^{*}}^{\alpha}\epsilon_{B^{*}}^{\beta},\\
\raisebox{-15pt}{\includegraphics[width=0.15%
\textwidth]{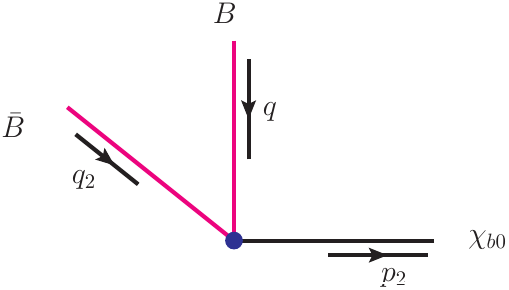}}           &\widehat{=}
&ig_{\chi_{b0}BB},\\
\raisebox{-15pt}{\includegraphics[width=0.15%
\textwidth]{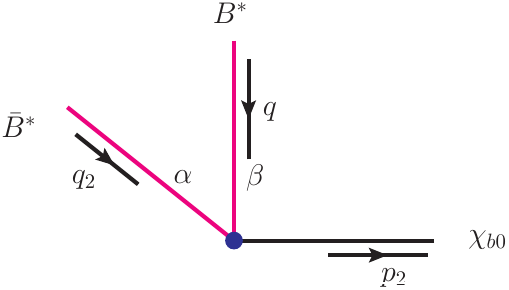}}       &\widehat{=}
&-ig_{\chi_{b0}B^*B^*}\epsilon_{\bar{B}^{*}}^{\alpha}\epsilon_{B^{*}}^{\beta}g_{\alpha\beta},\\
\raisebox{-15pt}{\includegraphics[width=0.15%
\textwidth]{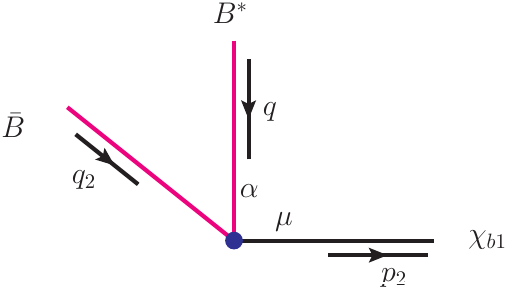}}           &\widehat{=}
&-ig_{\chi_{b1}BB^{*}}\epsilon_{\chi_{b1}}^{*\mu}\epsilon_{B^{*}}^{\alpha}g_{\mu\alpha},\\
\raisebox{-15pt}{\includegraphics[width=0.15%
\textwidth]{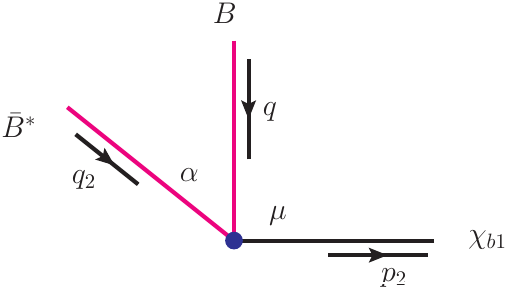}}       &\widehat{=}
&ig_{\chi_{b1}BB^{*}}\epsilon_{\chi_{b1}}^{*\mu}\epsilon_{\bar{B}^{*}}^{\alpha}g_{\mu\alpha},\\
\raisebox{-15pt}{\includegraphics[width=0.15%
\textwidth]{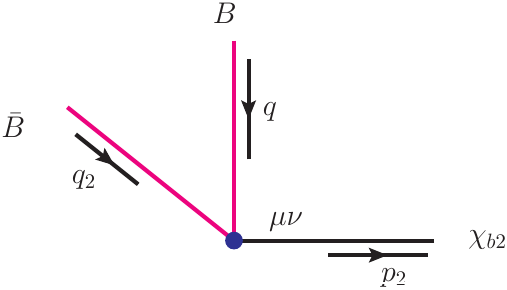}}           &\widehat{=}
&-ig_{\chi_{b2}BB}\epsilon_{\chi_{b2}}^{*\mu\nu}q_{\mu}q_{2\nu},\\
\raisebox{-15pt}{\includegraphics[width=0.15%
\textwidth]{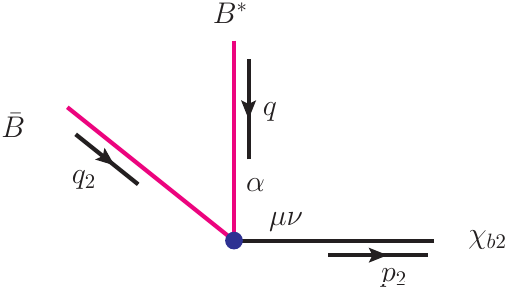}}       &\widehat{=}
&ig_{\chi_{b1}BB^*}\varepsilon_{\mu\rho\alpha\beta}\epsilon_{\chi_{b2}}^{*\mu\nu}p_{2}^{\rho}q_{\nu}q_{2}^{\beta}\epsilon_{B^{*}}^{\alpha},\\
\raisebox{-15pt}{\includegraphics[width=0.15%
\textwidth]{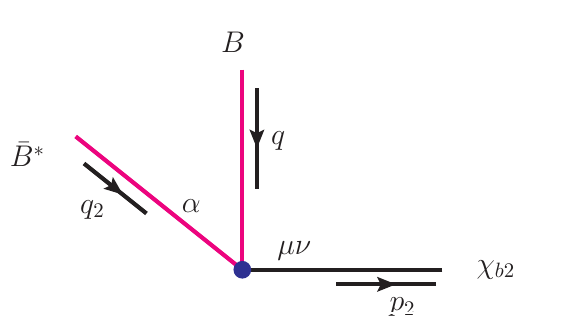}}       &\widehat{=}
&-ig_{\chi_{b1}BB^*}\varepsilon_{\mu\rho\alpha\beta}\epsilon_{\chi_{b2}}^{*\mu\nu}p_{2}^{\rho}q_{\nu}q_{2}^{\beta}\epsilon_{\bar{B}^{*}}^{\alpha},\\
\raisebox{-15pt}{\includegraphics[width=0.15%
\textwidth]{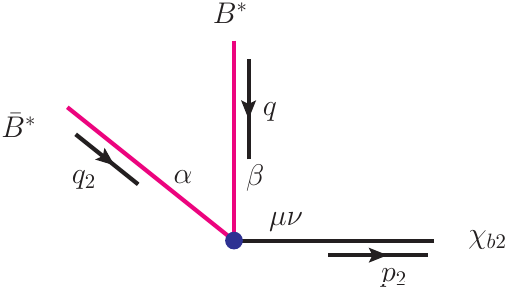}}   &\widehat{=}
&ig_{\chi_{b2}B^*B^*}\epsilon_{\chi_{b2}}^{*\mu\nu}\epsilon_{\bar{B}^{*}}^{\alpha}\epsilon_{B^{*}}^{\beta}g_{\mu\alpha}g_{\nu\beta},\\
\raisebox{-15pt}{\includegraphics[width=0.15%
\textwidth]{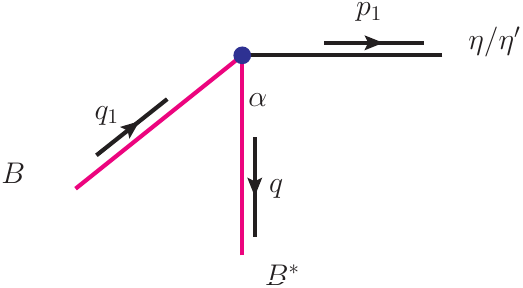}}           &\widehat{=}
&-g_{BB^*\eta^{(\prime)}}p_1^{\alpha}\epsilon_{B^{*}\alpha}^{*},\\
\raisebox{-15pt}{\includegraphics[width=0.15%
\textwidth]{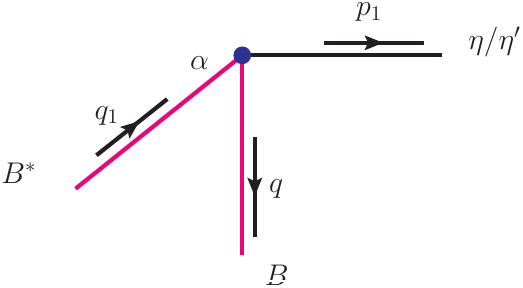}}       &\widehat{=}
&g_{BB^*\eta^{(\prime)}}p_1^{\alpha}\epsilon_{B^{*}\alpha},\\
\raisebox{-15pt}{\includegraphics[width=0.15%
\textwidth]{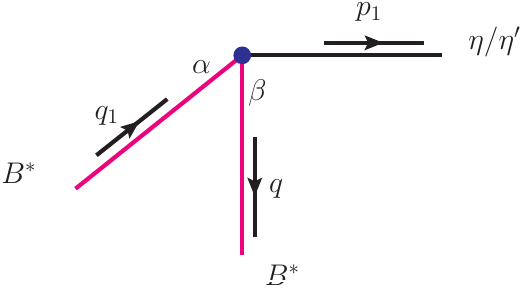}}       &\widehat{=}
&-g_{B^*B^*\eta^{(\prime)}}\varepsilon_{\mu\nu\alpha\beta}q_1^{\nu}q^{\mu}\epsilon_{B^{*}}^{\alpha}\epsilon_{B^{*}}^{*\beta},\\
\raisebox{-15pt}{\includegraphics[width=0.15%
\textwidth]{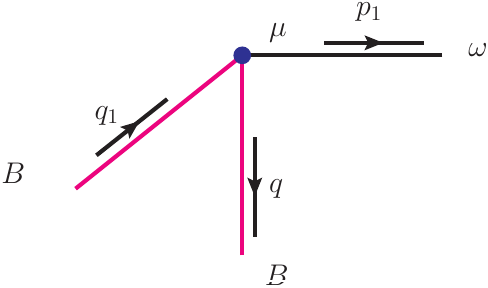}}           &\widehat{=}
&-g_{BB\omega}\epsilon_{\omega}^{*\mu}(q_{1\mu}+q_{\mu}),\\
\raisebox{-15pt}{\includegraphics[width=0.15%
\textwidth]{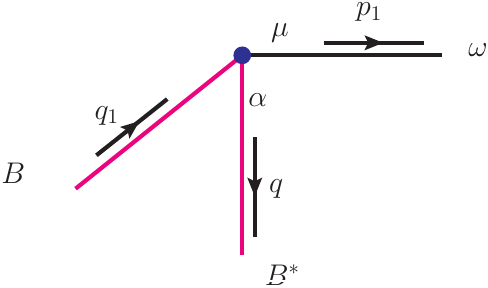}}       &\widehat{=}
&2f_{BB^*\omega}\varepsilon_{\mu\nu\alpha\beta}\epsilon_{\omega}^{*\mu}p_{1}^{\nu}(q_1^{\beta}-q^{\beta})\epsilon_{B^{*}}^{*\alpha},\\
\raisebox{-15pt}{\includegraphics[width=0.15%
\textwidth]{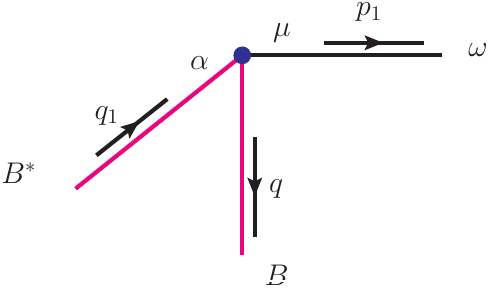}}       &\widehat{=}
&2f_{BB^*\omega}\varepsilon_{\mu\nu\alpha\beta}\epsilon_{\omega}^{*\mu}p_{1}^{\nu}(q^{\beta}-q_1^{\beta})\epsilon_{B^{*}}^{\alpha},\\
\raisebox{-15pt}{\includegraphics[width=0.15%
\textwidth]{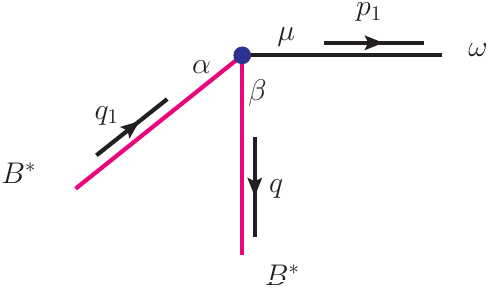}}   &\widehat{=}
&\begin{split}
&\epsilon_{\omega}^{*\mu}\left(g_{B^*B^*\omega}g_{\alpha\beta}(q_{1\mu}-q_{\mu})+4f_{B^*B^*\omega}\right.\\
&\left.\times(g_{\mu\beta}p_{1\alpha}-g_{\mu\alpha}p_{1\beta})\right)\epsilon_{B^{*}}^{\alpha}\epsilon_{B^{*}}^{*\beta}.
\end{split}
\end{eqnarray}

\end{document}